\documentclass[preprint]{aastex}

\usepackage{graphicx}
\usepackage{color}
\usepackage{amssymb}

\newcommand{\Me}{$M_{\earth}$}

\shorttitle{Volcanic Super-Earths}
\shortauthors{Kite, Manga and Gaidos}

\begin{document}

\title{Geodynamics and Rate of Volcanism on Massive Earth-like Planets}


\author{E.S. Kite\altaffilmark{1}}
\email{kite@berkeley.edu}

\author{M. Manga\altaffilmark{1}}
\affil{Department of Earth and Planetary Science, University of California at Berkeley, Berkeley, CA 94720}

\author{E. Gaidos}
\affil{Department of Geology and Geophysics, University of Hawaii at Manoa, Honolulu, HI 96822}


\altaffiltext{1}{Center for Integrative Planetary Science, University of California, Berkeley}
\begin{abstract}
We provide estimates of volcanism versus time for planets with Earth-like composition
and masses 0.25 - 25 {\Me}, as a step toward predicting atmospheric mass on extrasolar rocky planets.
Volcanism requires melting of the silicate mantle. We use a thermal evolution model, calibrated against Earth, in combination with
standard melting models, to explore the dependence of convection-driven decompression mantle melting on
planet mass. We show that (1) volcanism is likely to proceed on massive planets with plate tectonics over the main-sequence
lifetime of the parent star; (2) crustal thickness (and melting rate normalized to planet mass) is weakly dependent on planet mass; 
(3) stagnant lid planets live fast (they have higher rates of melting than their plate tectonic counterparts early in their thermal evolution), but die young (melting shuts down
after a few Gyr); (4) plate tectonics may not operate on high mass planets because of the production of buoyant crust which is difficult to subduct; and (5)
melting is
necessary but insufficient for efficient volcanic degassing -- volatiles partition into the earliest, deepest melts, which may be denser than the residue and sink
to the base of the mantle on young, massive planets.
Magma must also crystallize at or near the surface,
and the pressure of overlying volatiles must be fairly low, if volatiles are to reach the surface. If volcanism is detected in the $\tau$ Ceti system, and tidal forcing can be shown to be weak, this would be evidence for plate tectonics.
\end{abstract}




\section{Introduction}
Theory predicts the existence of rocky planets having 1-10 Earth masses (e.g. \citet{ida04}). Planets in this
mass range are now being detected \citep{riv05}, and next-decade observatories, such as \emph{James Webb Space Telescope} and \emph{Giant Magellan Telescope}, may be able to detect any atmospheres. 
A planet's atmosphere will consist of gas (1) accreted from the nebula, (2) degassed during impact accretion, and (3) degassed during subsequent geologic activity. (1) will depend on the lifetime of the nebula, (2) and (3) will depend on the volatile abundance of material \citep{elk08a}, and all will be modified by atmospheric escape. Very large planets far from their parent star will retain primitive gas, but smaller planets closer to their parent star will not.
Loss rates vary between gases, so planetary atmospheres could be a mixture of gases left over from the initial atmosphere,
and those replenished by volcanism. 
Thermal emission phase-curves gathered from extrasolar planets can set bounds on atmospheric
mass. Spectral observations
of atmospheric constituents with short photochemical lifetimes, such as SO$_2$,
 would require an ongoing source --– most
likely, volcanic degassing. 

Volcanism results from partial melting of upper mantle. (Planetary mantles can cool convectively without volcanism - present-day Mercury is almost certainly an example). 
Partial melting occurs when the adiabat crosses the solidus.
Assuming that the adiabat is steeper than the solidus, this requires that the potential temperature of the mantle $T_{p}$ exceed the zero--pressure
solidus of mantle rock (e.g., peridotite) (Figure \ref{PEDAGOG}a):
\begin{equation} T_p = T_{m,r} - P_{r}\frac{\partial V}{\partial S} \ge  T_{sol}(0), \end{equation}
where $T_{m,r}$ is the mantle temperature evaluated at some reference pressure $P_{r}$, \(\frac{\partial V}{\partial S}\) is the adiabat, potential temperature is defined as the temperature a parcel of solid
mantle would have if adiabatically lifted to the surface, and
$T_{sol}$ is the solidus, evaluated here at zero pressure.
On planets with plate tectonics, the thickness of the crust (the crystallized melt layer) is a convenient measure of the intensity of volcanism. The rate of
crust production is the product of crustal thickness, plate spreading rate, and mid-ocean ridge length.
The pressure at the base of the crust is the product of crustal thickness, the planet's surface gravity, and crustal density. A reasonable approximation to the pressure at the base of
the crust is the integral of the fractional-melting curve from great depth to
the surface (Figure \ref{PEDAGOG}a).

Venus and Mars lack plate tectonics: their mantles are capped by largely immobile,
so-called
`stagnant lid' lithospheres, which cool conductively. Because mantle cannot rise far into the stagnant lid,
melting can only occur if the temperature at the base of the stagnant lid exceeds the local solidus of
mantle rock:
\begin{equation} T_{m,r} - (P_{r} -P_{lith}) \frac{\partial V}{\partial S} \ge  T_{sol}( P_{lith} ), \end{equation}
where $P_{lith} = \rho_{lith} g Z_{lith}$ is the pressure at the base of the stagnant lid, $\rho _{lith}$ is lithospheric density, $g$ is gravity, and $Z_{lith}$ is stagnant lid thickness. This is a more stringent condition than (1)
if the adiabat is steeper than the solidus (Figure \ref{PEDAGOG}b). Io shows yet another style of rocky-planet mantle convection: magma pipe cooling. 

Previous studies have examined both rocky planet atmospheres and massive-earth geodynamics. \citet{elk08a} estimate the mass and composition of super-Earth atmospheres degassed during accretion. Complementary to that study, we emphasize long-term geological activity. There is disagreement over whether plate tectonics will operate on massive planets. One previous study uses scaling arguments to argue that higher gravity favors subduction \citep{val07}. Another study shows that subduction might never begin if the yield stress of old plate exceeds the stresses imposed by mantle convection, which fall steeply with increasing planet mass \citep{one07}. We do not consider yield stresses in this paper. Instead, we analyze five possible volcanism-related limits to plate tectonics, tracing the consequences in more detail than the short paper of \citep{val07}.
The approach of \citet{pap08} is most similar to that taken here. We differ from \citet{pap08} in that we neglect the pressure dependence of viscosity, use more realistic melting models, and account for energy advected by magma. 

Here we examine (1) the history of partial melting on planets with either plate tectonics or stagnant lid convection, and (2) the effect of melting and crust production on the stability of plate tectonics.
Our method is given in \S 2. We use three different melting models to compute the intensity of volcanism
for massive Earth-like planets of different ages. 
Our results are given in \S 3, for both plate tectonic (\S 3.2) and stagnant lid (\S 3.3) modes of mantle convection. We also trace the implications of galactic cosmochemical evolution for heat production and planetary thermal evolution (\S 3.4). We find that results differ greatly depending on the mode of convection. Massive Earths with plate tectonics will produce melt for at least as long as the age of the Galaxy, stagnant lid planets will not. 
In \S 4, we analyse the effect of melting on the style of mantle convection. We show that plate buoyancy is likely to be a severe problem, and may be limiting for plate tectonics.
In \S 5, we relate our results to atmospheric degassing, and discuss the possible suppression of degassing (and, perhaps,
melting) by the higher ocean pressures expected on massive Earth-like planets. Finally, in \S 6, we summarize our results; justify our approximations and model limitations; and compare our results to solar system data.

\section{Model description and inputs}


We use a model of internal structure (\S 2.1) to set boundary conditions for a simple model of mantle temperature evolution (\S 2.2), which in turn
forces a melting model (\S 2.3). 
Greenhouse-gas regulation of surface temperature could allow melting and degassing to feed back to mantle thermal evolution (e.g. \citet{len08}),
but we neglect this. Throughout, we assume whole-mantle convection. Rather than attempt to predict exoplanet properties solely from basic physics and chemistry, we tune our models
to reproduce the thickness of oceanic crust on present-day Earth.



\subsection{Radius and mantle depth}

Given our assumption of whole-mantle convection, we need to know only the mantle's outer and inner
radii. The crust is thin, so the top of the mantle is $\approx$ the planet's radius, ${R}$. 
\citet{val06} propose the scaling \(R/R_{\earth} = (M/M_{\earth} ) ^{\simeq 0.27} \). Here, we use instead the `modified polytrope' of \citet{sea07} (their
equation (23)) to set planet radius. However, we take \(R/R_{\earth} = (M/M_{\earth} ) ^{\sim0.25} \) in our scaling relationships (13) –-- (14) and (21). To find the Core-Mantle Boundary (CMB) radius for the \citet{sea07} scaling, we set mantle mass $M_{mantle} = 0.675 \hspace{0.1cm} M_{planet}$ and numerically integrate inward using a pure magnesioperovskite mantle composition,
 a 4th-order Burch-Murnaghan equation-of-state, and material properties from \citet{sea07}. 
Core-mantle boundary pressure is calculated to be
1.5 Mbar for 1 {\Me}, 
and 2.9 (6.9, 14, 40) Mbar for 2 (5, 10, 25) {\Me}.

\subsection{Thermal model}
For a convecting mantle with a mobile lithosphere, if heat is generated solely by mantle radioactivity and is equal to heat lost by cooling at the upper boundary layer, 

\begin{equation} Q = \frac{ M_{mantle} }{ A } \sum_{i=1}^4 H_0(i)e^{-\lambda_i t} = Nu \frac{ k ( T_m - T_s ) }{ d } \end{equation}
\begin{equation} Nu \approx \left( \frac{ g \alpha ( T_m - T_s ) d^3 }{ \kappa \nu (T) Ra _{cr} } \right) ^ \beta  \end{equation}
\begin{equation} \nu (T) = \nu_0 e^{ ( A_0 / T_m ) } = \nu_1 e^{ (- T_m / T_{\nu} ) }, \end{equation}
where $Q$ is lithospheric heat flux, $A$ is the planet's
surface area, $H$ is the radiogenic power per unit mass, $i$ = 1-4 are the principal long-lived radioisotopes ($^{40}$K, $^{232}$Th, 
$^{235}$U,
$^{238}$U), $\lambda$ is the decay constant, $t$ is time, $Nu$ (Nusselt number) is the dimensionless ratio of
total heat flow to conductive heat flow, $k$ is thermal conductivity, $T_{m}$ is mantle temperature, $T_{s}$ is surface
temperature, $d$ is the depth to the core-mantle boundary, $g$ is gravitational acceleration, $\alpha$ is thermal
expansivity, $\kappa$ is thermal diffusivity, $\nu$ is viscosity, $Ra_{cr}$ is the critical Raleigh number with value
$\sim$10$^{3}$, $\beta$ is 0.3 \citep{sch01}, $A_0$ is activation temperature, and $T_{\nu}$ is the temperature increase (decrease) that decreases (increases) viscosity by a factor of $e$ \citep{sch01}. Equation (5) contains two equivalent parameterized expressions for $T$. 
Values used for these parameters are given in (Table 2). 

We neglect the pressure dependence of viscosity \citep{pap08}, which cannot be fully captured by parameterized models. In effect, we assume that the viscosity beneath the upper boundary layer, rather than some volume- or mass-averaged mantle viscosity, determines the properties of the flow.


We use the canonical values for $H_{i}$ given by \citet{tur02}, who
estimate that 80\% of Earth's current mantle heat flux is supplied by radioactive decay. Although Earth's surface
heat flux is well constrained, the fraction of the flux out of the mantle that is due to radiogenic heat
production is not. Literature values vary from $\le$0.2 \citep{lyu07} to 0.8 \citep{tur02}, with low values increasingly favored \citep{loy07}. 
Variability in $H_{i}$ could swamp any size signal in rates of volcanism, an important uncertainty addressed in \S 3.4.
A useful rule of thumb is that a doubling a planet's concentration of radiogenic elements makes it behave like a planet with double the radius \citep{ste03}.

Equations (3) --– (5) can be solved directly for $T$. It is then easy to find the mass dependence of
temperature (\S 3.1; Figure \ref{UREYEQUALS1}). However, secular cooling significantly contributes to the heat flux at the
bottom of the lithosphere, so at a given time a planet will have a higher internal temperature and a higher heat flow than
these thermal equilibrium calculations would suggest. Planets of different masses follow parallel cooling
tracks \citep{ste03}, and internal temperature is regulated by the
dependence of mantle viscosity on temperature \citep{toz70}. A very simple model for mantle thermal
evolution with temperature-dependent viscosity in plate tectonic mode is \citep{sch01}: -
\begin{equation} \frac{\partial T}{\partial t} = \frac{H}{c} - k_1 ( T_m - T_s ) ^{ ( 1+\beta ) } exp \left( \frac{ - \beta A_0 }{ T_m }, \right)
\end{equation}
where
\begin{equation} k_1 = \frac{ A k }{ c d M_{mantle} } \left( \frac{ \alpha g d^3 }{ \kappa \nu _0 Ra_{cr} } \right) \end{equation}
and $c$ is the specific heat capacity of mantle rock. 

We integrate this model
forward in time from a hot start, using a fourth-order Runge-Kutta scheme.
Because of the exponential temperature dependence of convective velocity, the transient associated with the initial conditions decays on a 100 Myr timescale, provided that the planet has a `hot start'.
Hot starts are overwhelmingly likely for differentiated massive rocky planets. Our initial condition is $T_{m}$ = 3273 K, but our results are insensitive to increases in this value. 

For planets in stagnant lid mode, we use the scaling of
\citet{gra98}: -
\begin{equation} T_c = T_m - 2.23 \frac{ T_m^2 } { A_o }, \end{equation}
where $T_{c}$ is the temperature at the base of the stagnant lid; plausible values lead to \( (T_{m} - T_{c}) \ll T_{m} \).
Then:
\begin{equation} Nu \approx \left( \frac{ g \alpha ( T_m - T_c ) d^3 }{ \kappa \nu (T) Ra _{cr} } \right) ^ \beta.  \end{equation}
Since \( T_c > T_s \), stagnant lid convection is less efficient at transporting heat than plate tectonics.

\subsection{Melting model}


Earth generates 34 km$^3$ yr$^{-1}$ crust of which 63\% is by isoentropic decompression melting at
mid-ocean ridges \citep{bes01}. (This percentage understates the contribution of mid-ocean ridge melting to overall mantle degassing, because most of the volatile
flux at arcs is just recycled from subducting crust). After four decades of intensive study, this is also the
best-understood melting process \citep{jut99}. Beneath mid-ocean ridges, the mantle undergoes corner flow.
Melt is generated in a prism with triangular cross-section, ascends buoyantly, and is focused to a
narrow magma lens beneath the ridge. 
Petrological systematics require \citep{lan92}, and most melting models assume \citep{ghi02}, that the source magmas for mid-ocean ridge basalt
 melt fractionally or with small residual porosity, separate quickly, and suffer relatively
little re-equilibration during ascent. For more massive planets, these remain robust assumptions.
Buoyancy forces driving segregation are stronger and, because the pressure at which the solidus and
adiabat intersect is at a shallower absolute depth, the ascent pathways are shorter. Because of these attractive simplifications
and because mid-ocean ridge melting dominates Earth's crust production budget,
we focus on mid-ocean ridge melting in this paper.

Isoentropic decompression melting pathways are distinguished by their values
of potential temperature, $T_p$. 
Actual temperatures of near-surface magmas
are lower because of the latent heat of melting, the greater compressibility of melts with respect to
solids, and, usually less important, near-surface conductive cooling. All  mid-ocean-ridge melting
schemes are very sensitive to $T_p$, especially just above the zero-pressure solidus. That is because increasing $T_p$ both increases the pressure at which melting first occurs (lengthening the `melting
column') and also increases the mass fraction of melting ($X$) suffered by the top of the melt column (Figure \ref{PEDAGOG}c). 

With the above assumptions,
\begin{equation} P_{crust} = - \int_{P_o}^{P_f} X(T,P) dP \end{equation}
\begin{equation}T(P) = T \left( P+ \delta P \right)-\left( \frac{ \partial V }{ \partial S } \right) \delta P +\left( \frac{ \partial X }{ \partial P } \right) L \end{equation}
\begin{equation} T(P_o) = T_p + P_o \frac{ \partial V }{ \partial S }, \end{equation}
where $P_f$ = 0 in the case of plate tectonics or $P_f = P_{lith}$ (the pressure at the base of the lithosphere) in the
case of stagnant lid convection; $P_o$ is given by the intersection of the adiabat with the solidus, which is
the point on the adiabat $ \left( \frac{ \partial V}{ \partial S} \right) $ where $X$ = 0; $L$ is the latent heat of melting; and $S$ is entropy.

We use three models for $X(T,P)$. In order of increasing complexity, they are those of \citet{mck84} as extended in \citet{mck88} (henceforth
MB88), \citet{kat03} (henceforth K03), and (for plate tectonic models only) \citet{ghi02}, with the \citet{smi05} front-end (henceforth pMELTS).
MB88 and K03 are similar in that they fit simple functional forms to experimental data, with MB88 more widely used although it is constrained by fewer data.
pMELTS is a state-of-the-art model of phase equilibria for compositions similar to Earth's
mantle. We use pMELTS throughout the predicted melting range, even though the
model is only calibrated for use in the range 1–-3 GPa. For pMELTS, we assume continuous melting with a
residual porosity of 0.5\% (that is, melt fractions greater than 0.5\% are evacuated from the melting zone), 
and we use the mantle composition inferred to underly Earth's mid-ocean
ridge system \citep{wor05} with 500 ppm water. Representative results with all these models are shown in Figure \ref{CT}.

Each model is required to produce 7 km thick basaltic crust, which is the observed
value on Earth \citep{whi01}, after 4.5 Gyr on an Earth-mass planet undergoing plate tectonics.
We adjust the offset between potential temperature in the melting models, and the characteristic mantle
temperature used in the thermal model, to obtain the observed crustal thickness. 
The required offset $T_m - T_p$ is 741 K, 707 K, and 642 K, for the MB88, K03, and pMELTS models, respectively (Figure \ref{CT}).

\section{Model output}


\subsection{Simple Scaling Laws from Thermal Equilibrium Calculations}
The simplest possible rocky planet model assumes that the ratio of radiogenic heat production to
lithospheric heat flux (the convective Urey number), $Ur$ = 1. 
We take the radiogenic-element concentrations given by
\citet{tur02} (Table 1) and use (3) -- (5) to set mantle temperature. This is greater for more massive planets
because their decreased surface area/volume ratio requires higher heat fluxes (\& more vigorous
convection) to dispose of the same heat flux. From equations (3)--–(5), but neglecting the dependence of $Q$ on $T_m$:
\begin{equation} \left( \frac{Ra}{Ra_{\earth}} \right) ^{\beta} = \left( \frac{M}{M_{\earth}} \right) \left( \frac{A}{A_{\earth}} \right) \left( \frac{d}{d_{\earth}} \right) \end{equation}
which with $R \propto  M^{0.25 –- 0.28}$ \citep{val06} gives $Ra \propto M^{\approx 2.45}$; using this
simplification leads to a relation between temperature and mass
\begin{equation} \nu(T) \propto M^{-5/4} \end{equation}
--- which when inserted into Equation (5) gives a good fit to the results shown in Figure \ref{UREYEQUALS1}. Similar scaling
arguments with $R \propto M^{1/3}$ give $\nu(T) \propto M^{-8/9}$, i.e. an approximately straight line on a linear-log graph of temperature versus mass
(Figure \ref{UREYEQUALS1}).




\subsection{Plate tectonics}
We now turn to our time-dependent results. 

\begin{list}{$\bullet$}
{ \setlength{\labelwidth}{2cm} }
\item \emph{Thermal evolution.} As anticipated \citep{ste03}, mantle temperatures for planets of different masses follow
$\sim$parallel cooling curves. Planets with $M$ = 2 (5, 10, 25) {\Me} have potential temperatures 39 K (97 K, 146 K, 221 K) greater than Earth after 4.5 Gyr (Figure
\ref{MASS}). There is little difference in thermal evolution between internal structure models: the temperature difference between
\citet{val06} and \citet{sea07} is always $<$15 K (Figure \ref{UREYEQUALS1}). The constant-density planet runs
significantly (up to 100K) colder since it has a much larger surface area, but the same radiogenic-element complement. From now on we use only the thermal evolution calculations for the \citet{sea07} internal-structure model.

\item \emph{Simple melting models, MB88 and K03.} Potential temperature increases
monotonically with mass, so the pressure at the base of the crust also increases monotonically (Figure \ref{CT}).
However, the absolute thickness of the crust also scales as the inverse of gravity. In other words, although bigger planets
run hotter, higher surface gravity moves the solidus and suppresses melting. For temperatures
close to the solidus, the first effect dominates, and increasing planet mass increases crustal thickness
(Figure \ref{CRUSTVSTIME}b \& c). Young and/or large planets show the opposite trend, with crustal thickness
decreasing as planet mass increases. Crustal thicknesses are within a factor of two of each other for 1 -- 25 {\Me} until 8.6 Gyr. After that, the ratio of crustal thicknesses diverges, as melting begins to
shut down on the lowest-mass planets. In both models, for planets of intermediate mass and with ages slightly greater
than the Solar System, increasing mass has only a small (and negative) effect on crustal thickness.
However, crust production per unit time increases with increasing mass, because more massive planets
have more rapid plate spreading: see \S 5.1. 

\item \emph{pMELTS, a more complex melting model.} pMELTS predicts crustal thickness will increase rapidly with increasing planet mass
for massive planets with ages comparable to the Solar System (Figure \ref{CRUSTVSTIME}d). Potential temperatures for these planets
are $>$ 1500 $^{\circ}$C. The crustal-thickness result at this temperature should be treated with suspicion, because when
$T_p$ $>$ 1460 $^{\circ}$C, pMELTS predicts melting will occur to pressures greater than those for which it has been
experimentally calibrated ($<$ 3 GPa). Crustal thickness is $>$ 1 km even after 13 Gyr in the pMELTs model,
even for masses of 1 {\Me}. This is because pMELTS-predicted crustal thickness ramps up slowly to 7 km crustal thickness as
temperature increases (Figure \ref{CT}), so contours of constant crustal thickness are spaced more widely in temperature (equivalently,
time) than with the other models. As with the other melting models, the planet mass that produces the thickest crust at a given time increases as the planets age.

 \end{list}

To summarize, mass dependence increases with time as planets cool toward the solidus, and (for a given mass range) the
sign of mass dependence changes with time. This is because of the strongly nonlinear behaviour of
melt production near the solidus; melting goes from zero to significant over a small range in potential
temperature. Low-mass planets approach this temperature by 7-8 Gyr. Their crustal thickness declines
more rapidly than on high-mass planets.

\subsection{Stagnant lid}

\begin{list}{$\bullet$}
{ \setlength{\labelwidth}{2cm} }
\item \emph{Thermal evolution.} Because stagnant lid convection is less efficient at transferring heat than plate tectonics, a planet in which
plate tectonics is suddenly halted will heat up (the thin solid line in Figure \ref{MANTLEMODE}).
This temperature rise reduces mantle viscosity, so $Ra$ increases. Temperature converges on the
evolutionary track of a planet that has always been in stagnant lid mode, with a characteristic
convergence timescale of $(H \Delta T_{mode})/cM_{mantle} \sim$ 1 Ga. A similar argument explains the temperature changes
associated with going from stagnant lid mode to plate tectonics. The temperature difference between
the tracks is $\Delta T_{mode} \approx$ 160 K for all masses. This is roughly \(T_{0}\ ln \left( ( T_{m} - T_{s} ) / ( T_{m} - T_{c} ) \right)  \), which can be understood by 
equating the RHS of Equations (4) and (9). Therefore, the thermal evolution of stagnant lid planets follows Figure \ref{CRUSTVSTIME}a, but $\approx$ 160K hotter.

\item \emph{Stagnant lid melting -- live fast and die young.} Melt production within an ascending column of mantle in stagnant lid mode is
truncated at the base of the lithosphere. For the same mantle temperatures, predicted erupted thickness is much smaller (Equation 10).
This effect opposes the increased temperature of stagnant-lid mantles.
To produce Figure \ref{CRUSTVSTIMESL} we use the same mantle-potential temperature offsets as in plate
tectonic mode (so the model is still `tuned to Earth'). Most planets run hotter in stagnant-lid mode to the extent that pMELTS
cannot be used, as $T_p$ exceeds the range over which it is calibrated. The MB88 and K03 models show roughly the same behavior in stagnant lid mode (Figure \ref{CRUSTVSTIMESL}). For young ($<$2-3 Gyr) planets, an
ascending column of mantle produces more melt in a stagnant lid mode than in plate tectonic mode –--
the higher temperature matters more than the (small) lithospheric thickness. For somewhat older
planets, an ascending column of mantle produces less melt in stagnant lid mode than in plate tectonic
mode. The temperature difference is much the same, but the growing lithosphere increasingly truncates the melting column. At a mass-dependent age
much less than the age of the Galaxy, melting ceases.
 
 \item \emph{What controls cessation of melting?} For a given melt production function, all planets in plate tectonics mode will cease volcanism at the
same potential temperature. Consequently, a planet's volcanic lifetime is delineated by an isotherm ($T_p$ = 1080 -– 1193 $^{\circ}$C,
depending on the melting model). For planets in stagnant lid mode, this is not the case. There is still a one-to-one
relationship between temperature and the absolute thickness of the lithosphere. However, the
absolute thickness of the melt zone at fixed temperature decreases with increasing gravity, but $z_{lith}$ does not. Thus, $z_{lith}$ increasingly truncates the melt zone as gravity increases. As a result, there is a temperature range for which low-mass planets can sustain melting in stagnant lid mode, whereas high--mass
planets cannot. Over the mass range 1 -- 25 {\Me}, this temperature range is $\sim$ 180K. Consequently, in
stagnant lid mode, more massive planets run much hotter but cease melting only moderately later than
smaller planets (Figure \ref{CRUSTVSTIMESL}). 
 \end{list}

\subsection{Initial bulk-chemistry and initial radiogenic-power variations}

`Non-canonical' initial radiogenic element complements have been suggested for Earth.
To evaluate this possibility, we show 1 \Me \hspace{0.03 cm} thermal evolution tracks with radioisotope complements appropriate to `undepleted' Earth \citep{rin91}, CI chondrites \citep{and89}, and EH chondrites \citep{new95} (Table 1; Figure \ref{H}). The CI and `undepleted'  tracks show similar behaviour to T\&S, but the U$^{235}$-rich, U$^{238}$-poor EH chondrite track shows more rapid cooling.
Since we use Earth's observed oceanic crust thickness to tune mantle
temperature, it is not particularly important to get the absolute values right, and three of the four radiogenic-element complements tested have similar behaviour over geological time.

The long-term thermal evolution of rocky planets depends on the
abundance of the long-lived radiosotopes $^{232}$Th,
$^{235}$U, and $^{238}$U at the time of planet formation. These are produced only by the rapid neutron
capture process ($r$-process) acting on the iron-peak isotopes.  This
is thought to occur only during explosive nucleosynthesis in stars
with $10-20 M_{\odot}$ \citep{Chen06}.  In contrast, Si is produced
during $\alpha$-chain process by the whole range of massive stars.
Th, and especially U, are difficult to detect in stars but europium
(Eu), another exclusively r-process element, can be readily measured.
The average observed stellar abundance of Eu to silicon decreases by a
factor of 0.63 as the abundance of heavy elements or metallicity
(represented by iron Fe) increases by a factor of 100 to the solar
value \citep{Cescutti08}.  The $r$-process appears to be universal and
all $r$-process elements scale closely with solar values
\citep{Frebel08}.  Therefore the average abundance of $^{232}$Th,
$^{235}$U and $^{238}$U isotopes can be predicted using the
trend of Eu with abundance or metallicity, the age-metallicity
relationship of the Galaxy, the star formation history of the Galaxy,
and the half life of each isotope.  We adopt a simple linear
age-metallicity relationship with an increase of 1 dex (factor of 2.5)
over the age of the Galaxy, with solar metallicity occuring 4.6 Gyr
ago (e.g. \citet{Pont04}).  Figure
\ref{ISOTOP} plots the predicted abundance of the three isotopes
using the observed trend of Eu and the Prantzos \& Silk (1998)
parameterization of the star formation history of the Galaxy.  (The
predictions are only weakly sensitive to the model of star formation).
The age of the Galaxy is taken to be 13.6 Gyr.  All abundances are
normalized to the value at the formation of the Sun.

Planets forming early in the history of the Galaxy would have 50\%
more $^{238}$U, but 6 times more $^{235}$U, than Earth.  The higher abundance is
because the amount of radioisotopes in the interstellar medium only
reflects massive star formation over a few half-lives, whereas $^{28}$Si
and other stable isotopes accumulate over the history of the Galaxy.
Therefore these systems are not U- and Th-rich, they are Si-poor.  The
high abundance of $^{235}$U could have an important role in the \emph{early}
thermal history of such planets.

The effect of these trends on \emph{present-day} planet temperatures, while still significant, is more modest. Figure \ref{GCE}
plots planet temperature against the age of the host star. Comparison with Figure \ref{CRUSTVSTIME}a shows that inclusion
of cosmochemical trends in $H_{i}$ lowers $T_{m}$ by up to 50 K for young planets, while raising $T_{m}$ by up to
40 K for old stars, compared to their present-day temperature had they formed with an Earthlike inventory of radiogenic elements.

We have assumed that the major-element composition of planetary mantles is similar everywhere and at all times. This is unlikely to be true in detail: for example, it has
been proposed that on early Earth the mid-ocean-ridge-basalt source was more depleted than at the present day \citep{dav07b}. Earth's
continent mass fraction may be higher \citep{ros06}, or lower than is typical.
More severe variations in major-element composition,
with correspondingly major shifts in rheology and in the solidus, can be imagined (e.g., \citet{gai00, kuc05}). Even highly oxidized `coreless' planets have been modelled \citep{elk08b}.
We leave the geodynamic consequences of such variations as an open subject: a future objective will be to use geodynamic observables to constrain internal structure and bulk composition.


\section{Effect of Melting on Style of Mantle Convection}
We have shown that transitions between plate tectonics and stagnant lid mode have an impact on thermal evolution
comparable to that of mass over the range of masses considered in this paper. 
With this motivation, we now assess whether plate tectonics is viable on more massive Earth-like
planets. Our approach will be to use our comparatively robust understanding of melting to examine conditions under which other forms of heat transfer supplant Earthlike plate tectonics. Vertical tectonics, as seen on Io (\citet{moo03,lop07}) may be thought to
take over from horizontal (plate) tectonics when either:-- (1) the thickness of
the crust becomes comparable to that of the lithosphere; (2) heat lost by magma transport dominates over
conduction, or (3) the crust delaminates. We also assess the likelihood that (4) continental growth or (5) greater plate
buoyancy prevents subduction. 
We present results only for our fiducial calculation with MB88 melting and radiogenic-isotope complements following \citet{tur02}.

We assume throughout that surface water is available to hydrate lithosphere rock. Weakening the lithosphere by hydration is thought to be a prerequisite for plate tectonics.

\subsection{Crust thicker than lithosphere}
If the crustal thickness $Z_{crust}$ is comparable to the lithospheric thickness $Z_{lith}$, the lower crust is likely to
melt and form buoyant diapirs. Widespread intracrustal diapirism within the oceanic
crust is not known on Earth and, if it were a major heat sink for the mantle, would be a substitute for
plate tectonics. $Z_{lith}$ scales as $Q^{-1}$, and on Earth the equilibrium value of $Z_{lith}$ is $\sim$ 110 km
\citep{mck05}. Therefore,

\begin{equation} \left( \frac{ Z_{crust} }{ Z_{lith} } \right) = \frac{ 7 }{ 110 } \left( \frac{ Q }{ Q_{Earth} } \right) \left( \frac{ Z_{crust} }{ Z_{crust,Earth} } \right). \end{equation}

We find $Z_{crust}/Z_{lith}$ $<$ 1 for all planets $>$ 2 Gya (Figure \ref{COVERL}), so intracrustal diapirism is unlikely within equilibrium
lithosphere. 
Intracrustal diapirism is unlikely to be the limiting
factor for super-Earth plate tectonics.
%


\subsection{Magma-pipe transport energetically trumps conduction}
On Earth, heat lost by conduction through thin lithosphere near mid-ocean ridges greatly exceeds heat
lost by advection of magma. On Io, the opposite is true: most internally-deposited heat is lost by
advection of magma through lithosphere-crossing magma conduits -- `magma pipes' \citep{moo01}. Such a planet, although it may have limited plate spreading, is not in plate tectonic
mode; vertical rather than horizontal motion of the material making up the lid is the more important process.

We introduce the dimensionless `Moore number', $Mo$, in appreciation of the work of \citet{moo01,moo03}, 
 which we define to be the ratio
of magma-pipe heat transport, $Q_{magma}$, to heat lost by conduction across the lithosphere, $Q_{cond}$.
To calculate how this number (analogous to a 
Peclet number) scales with increased mass, we specify that total heat loss (magma pipe
transport plus conduction across the boundary layer) adjusts to match the heat loss across the boundary
layer prescribed by the thermal evolution model:

\begin{equation} Mo = \frac{ Q_{magma} }{ Q_{cond} } = \frac{Q}{Q_{cond}} - 1 \end{equation}

An upper bound on magma-pipe transport is to assume that all melt crystallizes completely and cools to the surface temperature.
In that case:

\begin{equation} Q_{magma} =  \rho_{crust} ( c_{b} \Delta T_{1} + E_{lc} ) Z_{crust} l_{r} s \sim c_{1} s. \end{equation}

where $\rho_{crust}$ is crustal density, $c_b$ is the specific heat of basalt, $\Delta T_1$ is the temperature contrast between lava and the surface, $E_{lc}$ is the latent heat of crystallization, $l_r$ is ridge length and $s$ is spreading rate. Provided that half-space cooling is a good approximation to the thermal evolution of plates,

\begin{equation} Q_{cond} =   2k \Delta T_{2} \left( \sqrt{ \frac{ A_{oc} l_{r} }{ \pi \kappa } } \right) \sqrt{ s } \sim c_{2} \sqrt{ s }, \end{equation}

where $A_{oc}$ is the area of the ocean basins only, which we take to be 0.6 $\times$ $A$ as on Earth. We can now solve the quadratic in $s$ ---

\begin{equation} Q^2 =  c_{1}^2 s^2 + c_{2}^2 s.  \end{equation}

Finally we obtain an expression for $Mo$:

\begin{equation} Mo = \frac{ Q_{magma} }{ Q_{cond} } = \left( \frac{ \rho_{crust} (c_{b} \Delta T_{1} + E_{lc} ) Z_{crust} l_{r} }{ 2k \Delta T_{2}
\sqrt{ A l_{r} } / ( \sqrt{ \pi \kappa } ) } \right) \sqrt s. \end{equation}

Here $l_r$ is ridge length and $s$ is spreading rate. We assume a latent heat of crystallisation $E_{lc}$ = 550 kJ/kg,
a specific heat of basalt $c_b$ = 0.84 kJ/kg/K, crustal density $\rho_{crust}$ of 2860 kg/m$^3$ \citep{car90},
base lithosphere temperature of 1300 $^{\circ}$C, and lava temperature of 1100 $^{\circ}$C and surface temperature of 0 $^{\circ}$C
giving $\Delta T_1$ = 1100 K and $\Delta T_2$ = 1300K. This gives 4.2 x $10^{18}$ J in melt/km$^3$ plate made. As an illustration,
for Earth (mid-ocean-ridge crust production 21 km$^3$ yr$^{-1}$), $Q_{magma}$ is 2.8 TW. The total oceanic heat flux is 32 TW,
so today's Earth has a $Mo \sim$ 0.10.

 Therefore, given the one-to-one relationship between $Q$ and $T_p$, we can solve for
spreading rate (Figure \ref{SPREADING}). As $Mo \rightarrow 0$, $s \rightarrow Q^2$ \citep{sle01}. Notice that the characteristic age at subduction $\tau = A / (2 l_r s)$.  As a consequence, the Moore number does not vary with different possible scalings of ridge length (equivalently, plate size) with increased planet mass.
As planet mass increases, we hold plate area (rather than the number of plates) constant to produce Figure \ref{SPREADING}. If one instead holds the number of plates constant, $l_r$ falls and $s$ increases, but, because $\tau$ is unchanged, our subsequent buoyancy and rate of volcanism calculations (\S 4.4 \& \S 5) are not affected. 
For Earth, (18) gives $s$ $\sim$ 5 cm $yr^{-1}$, a good match to present-day observations (minimum  $<$ 1 cm $yr^{-1}$ in the Arctic Ocean, maximum 15 cm $yr^{-1}$ at the East Pacific Rise, mean $s$ $\sim$ 4 cm $yr^{-1}$).

Reaching $Mo \sim$ 1 requires temperatures near or beyond the limit of validity of our melting models.
Coincidentally, $Mo$ = 1 plots close to $Z_{crust}/Z_{lith}$ = 1 on a mass-time graph. We conclude that heat-pipe cooling is not dominant for Earth-like planets $>$ 2 Gyr old.

\subsection{Phase transitions within crust}
Basalt, which is less dense than mantle rock, undergoes a high-pressure exothermic phase transition to
eclogite, which is more dense than mantle rock \citep{buc02}. Except during subduction, the
pressures at the base of Earth's basaltic crust are too low to make eclogite. However, the pressures at
the base of the crust on massive Earth-like planets are far greater. If the crust of a massive earth-like
planet includes an eclogite layer at its base, this may delaminate and founder, being refreshed by hotter
mantle \citep{vla94}.

We tracked temperature and pressure at the base of the crust for planets in plate tectonics mode, and
compared these with the phase boundaries for eclogite plotted in Fig. 9.9 of \citet{buc02}.
We assumed uniform thermal conductivity within the thermal boundary layer. We found that eclogite is
not stable for $M$ $<$ 25 {\Me}. As planet mass increases, the ratio of crustal to lithospheric thickness
increases, while mantle temperature also rises. Consequently, the temperature at the base of the crust
rises, inhibiting the exothermic eclogite-forming reaction. 

\subsection{A continental throttle?}
Continental crust may sequester radiogenic elements, inhibiting plate tectonics and melting by
imposing more rapid mantle cooling. The limit is a fully differentiated planet, in which successive
melting events have distilled nearly all radiogenic elements into a thin shell near the surface. Independent of
this effect, too great an area of nonsubductible, insulating continents may itself
be enough to choke off plate tectonics –- a limit of 50\% of total area has been suggested \citep{len05}. Which limit is more
restrictive to plate tectonics? On Earth, continents contain (26-77)\% of the radiogenic complement of
Bulk Silicate Earth \citep{kor08}, and cover $\sim$ 40\% of the planet's surface area. If the threshold
value for a significant nonsubductible-cover effect on mantle dynamics is 50\% \citep{len05},
this limit will be passed before complete differentiation occurs. However, all radiogenic elements
would be sequestered in continental crust before continental coverage reached 100\%. 
We expect that the thickness contrast between
continental and oceanic crust will scale as the inverse of gravity. 
This is because crustal thickness variations are limited by crustal flow (and brittle failure down gravitational-potential-energy gradients).
If in addition the radiogenic-element
content of continental crust does not vary, then we can relate the two limits by expressing the fraction of
planet surface area covered by continents as

\begin{equation} f_{area} = \frac{ V_{cont} }{ A Z_{crust}} \propto \frac{ f_{radio} M}{ M^{1/2} M^{-1/2} }\propto f_{radio} M.  \end{equation}

using the approximation $R \propto M^{\sim 0.25}$. This implies that continental coverage will be the more severe limit
for massive planets as well. A massive planet will enshroud itself with with nonsubductible crust before it sequesters
a substantial fraction of its radiogenic elements into crust.

Provided that crustal flow limits continental
thickness, a representative calculation shows that continents will spread out to coat the surface of
an Earth-like planet $>$ 3 {\Me} in much less than the age of the Earth (Figure \ref{CONTTHROTTLE}); this is because continental production rate
scales roughly as planet mass, but planet area increases only as the square root of mass. To produce this figure, we set net
continental growth to zero for the first 1 Gyr of each planet's evolution (guided by the age of the oldest surviving continental blocks on Earth). From 1 Gyr forward, we set continental growth to be
proportional to crustal production rate, with a proportionality constant picked to obtain 40\% coverage
on Earth today.


\subsection{Will trenches jam?}

Subduction will cease if the relative buoyancy of crust, less dense than mantle, exceeds that of the colder and denser lithospheric mantle.
Provided that thermal conductivity and crustal density are constant, the subduction condition is

\begin{equation}\Delta\rho \cong - \rho_{lith} + \frac{ 1 }{ Z_{lith}} \left( \rho_{lith} (1 + \alpha \Delta T_{3} ) ( Z_{lith} - Z_{crust} ) + ( \rho_{crust} Z_{crust} ) \right) < 0 \end{equation}

where

\begin{equation} \Delta T_{3} =  \frac{1}{2} \left(1 - \frac{ Z_{crust} } { Z_{lith}} \right) \Delta T_{2} \end{equation}

and $\Delta\rho$ is the density difference favoring subduction, $\rho_{lith}$ is the reference density of mantle underlying the plate, $Z_{lith}$ is lithospheric thickness, $\Delta T_{3}$ is the average cooling of mantle lithosphere, all evaluated at subduction.

Hotter –-- that is, bigger or younger --– planets must recycle plate faster, so a plate has less time to cool.
In addition, higher potential temperatures produce a thicker crust. Both factors tend to produce
positively buoyant plate, which is harder to subduct. This effect is more severe for massive planets
because of their greater gravity.

Once subduction is initiated the basalt-to-eclogite transition can sustain subduction.
However, it is not clear that subduction initiation is possible if plate is
positively buoyant everywhere, as seems likely for Early Earth \citep{dav92,sle00}. The geological
record is neither mute nor decisive. All but the last Gyr of Earth's tectonic history is disputed, but
evidence is accumulating that subduction first occurred at least 2.7 Gya (1.8 Ga after formation), and
perhaps earlier than 3.2 Gya (1.3 Ga after formation) \citep{con08}. Taking buoyancy stress per unit length of trench to be the appropriate metric for
buoyancy, we can relate Archean observations to high mass planets (Figure \ref{BUOYSTRESS}). For
example, the buoyancy stress that had to be overcome on Earth after 1.8 Ga is the same as that on a 16
{\Me} planet after 4.5 Ga. 
Here we have assumed
that all plate reaches the subduction zone at a characteristic age, $\tau$, that the temperature distribution
within the plate is described by half-space cooling (so $Z_{lith}$ = 2.32 $\kappa^{0.5} \tau^{0.5}$), and that $k$ and $\rho_{crust}$ (2860 kg
m$^{-3}$ ; \citet{car90}) are constant. This gives a negative (subduction-favoring) buoyancy stress of 38 MPa at the characteristic age of subduction on
 the present Earth. The much more sophisticated model of \citet{afo07} yields 21 MPa, so our estimate probably understates forces retarding subduction.
 However, the base of a thick, water-rich crust may be at temperatures/pressures permitting
amphibolite-grade metamorphism, producing dense crust. (In addition, 
the crust-mantle density contrast declines with increasing
melt fraction as the olivine content of the crust increases). For this reason, we also
plot results for a constant crustal density of 3000 kg m$^{-3}$, intermediate between the densities of
amphibolite (3000 –-- 3300 kg m$^{-3}$; \citet{clo93}) and unmetamorphosed crust. 
Whether or not amphibolitization is considered, our assessment is that plate
buoyancy raises a severe hurdle for plate tectonics on massive Earths, and may well be limiting.

\section{Rate of volcanism and implications for degassing}

\subsection{Degassing rate}
In plate tectonics mode, for `Earth-like' planets with oceans and some land, we calculate the
rate of volcanism by multiplying spreading rate (\S 4.3), mid-ocean ridge length, crustal thickness (\S 3), and crustal density,
then dividing by planet mass (Figure \ref{RATE}). Our model gives 1.2 x 10$^{-11}$ yr$^{-1}$ by mass for Earth, observations give 1.1 x 10$^{-11}$ yr$^{-1}$ \citep{bes01}.
The discrepancy is mainly due to our model's higher-than-observed spreading rate (\S 4.3).
For comparison, Earth's total observed present-day rate of volcanism is
1.7 x 10$^{-11}$ yr$^{-1}$ by mass, and 3.1 x 10$^{-11}$ yr$^{-1}$ by volume \citep{bes01}. This includes arc and ocean-island volcanoes, which are not modelled in this paper. Rate per
unit mass increases monotonically with increasing mass, for all times.
Per unit mass, rates of volcanism vary only by a factor of three on planets $<$ 3 Gyr
old and $>$ 1 {\Me}, but a stronger mass dependence develops for older planets (Figure \ref{RATE}). Melting in plate-tectonic
mode ceases when potential temperature falls below the zero-pressure solidus, but this only occurs for small (0.25 {\Me}) planets $\ge$ 10 Gya (Figure \ref{CRUSTVSTIME}b).

In plate tectonic mode, the flux of mantle into the melting zone balances spreading rate 
To calculate the rate of volcanism on a planet in stagnant lid mode (Figure \ref{RATESL}), we must first find
 the flux of material into the upper boundary layer as a function of $Ra$. We define a near-surface convective velocity such that all
 heat flow in excess of the conductive heat flow is advected: -

\begin{eqnarray*} Nu - 1 = \frac{u \Delta T_{conv} \rho c }{2 \left( \frac{k \Delta T_{cond}}{d} \right) } = \frac{u \rho c d}{2 k} \left( \frac{T_m - T_c}{T_m - T_s} \right) \end{eqnarray*}
\begin{equation} 
u = 2 (Nu - 1) \left( \frac{k}{\rho c d} \right) \left( \frac{T_m - T_s}{T_m - T_c} \right) 
\end{equation}


where the factor of 2 takes account of the need for downwelling material to balance upwelling. The crust production mass flux, normalized to planet mass, is

\begin{equation} R = \frac{A u \rho}{ M_{planet} } \left( \frac{  \rho_{crust} Z_{crust} g }{P_f - P_o} \right), \end{equation}

where $R$ is the rate of melt generation per unit mass, $Z_{crust}$ is obtained from our models in \S3, and the bracketed term is the mean melt fraction.
This sets an upper limit to the rate of volcanism (Figure \ref{RATESL}) because it assumes that all melt generated reaches the surface,
and that all ascending mantle parcels reach $P_o$.
The first assumption is probably safe because of the large forces driving magma ascent on massive planets,
and the short absolute distances they must ascend. With these assumptions, we find that the rate of volcanism on stagnant lid planets 
initially exceeds that on their plate tectonic counterparts 
but this contrast soon reverses as the stagnant lid thickens. The predicted shutdown
of non-plume melting on stagnant lid planets 
shows remarkably weak dependency on mass and melting model, as explained in \S 3.3. 

We consider the rate of crust production to be a good proxy for the rate of degassing. 
 But because volatiles are incompatible,
they partition almost quantitatively into even small fractions of melt. Therefore, a more accurate statement is that
degassing should be proportional to the flux of mantle processed through melting zones \citep{pap08}. We had great difficulty in determining this processing flux because of the `solidus rollover' problem (\S 6.2). 
(Because of volatile incompatability, order-unity differences in the
volatile concentrations of planetary mantles should not make much of a difference to our melting curves. Wet parcels of ascending mantle
soon `dry out'.) In addition to lavas that are extruded at the surface,
magmas that crystallize below the surface as intrusions -- such as the
sheeted dykes and gabbros of Earth's oceanic crust -- are assumed to degas fully and are referred to as
`volcanism'. 

The implications for greenhouse-gas release of our results depend on the water and CO$_2$ contents of extrasolar planet mantles.
These will depend on the mantle's oxidation state \citep{elk08b}, the range of semimajor axes from which the growing planet draws material,
and the extent of volatile loss on the planetesimals which collide to form the planet. The water and CO$_2$ contents of terrestrial magmas are our only empirical guide.
At mid-ocean ridges, these range from 0.1\% -- 1.5\% for H$_2$O,
and 50-400 ppm for CO$_2$ \citep{opp03}. Taking the upper-limit values
(which is appropriate, given that additional volatiles are almost certainly present in an
unmeasured fluid phase),
assuming that all mantle fluxing through the melting zone is completely degassed,
and ignoring overburden pressure,
our model yields 6 x $10^{13}$ mol a$^{-1}$ for H$_2$O and 7 x $10^{11}$ mol a$^{-1}$ for CO$_2$.
This is within the range of uncertainty of observational estimates. Because our model is not `tuned' to observed rates of volcanism, but only to crustal thickness,
 this lends some credence to our results for other Earth-like planets. 
 
\subsection{Overburden pressure}
The rate of degassing from seafloor volcanoes
will also be regulated by the thickness of the volatile overburden. A volatile envelope can have two effects: --

(1) through a greenhouse effect, a volatile envelope can raise surface temperatures, and increase partial melting. 
This is shown for our thermal-equilibrium model in Figure \ref{UREYEQUALS1}.
Here, we show the internal temperature needed to drive convection when the
surface temperature is 647K, the critical point of water (q.v. 273 K for the baseline model). In thermal
equilibrium, the reduced mantle-surface temperature difference demands more vigorous convection to
drive the same heat flux across the upper boundary layer. The increase in mantle temperature is 50-55K
(Figure \ref{UREYEQUALS1}), which has a significant feedback effect on melting (Figure \ref{CT}). Such high
surface temperatures are classically considered to be encountered only for brief intervals on the path to
a runaway greenhouse (Ingersoll, 1969), but if this catastrophe is suppressed for massive Earths
(Pierrehumbert, 2007), massive Earths will experience these temperatures for geologically significant
intervals. The sign of this feedback is positive; higher mantle temperatures increase crustal thickness,
and the associated degassing would enhance the greenhouse effect. We do not consider still higher temperatures, 
relevant for close-in rocky exoplanets \citep{gai07}, because
plate tectonics is thought to require liquid water.

(2) through overburden pressure, a volatile envelope can suppress degassing (and melting for sufficiently thick volatile layers).
Water degassing is readily suppressed by 0.1-0.2 GPa of
overburden \citep{pap97}, although joint-solubility effects allow some water to degas at higher pressures in association 
with other gases \citep{pap99}. CO$_{2}$, an important regulator of climate on the known
terrestrial planets, has a solubility of ~ 0.005\% GPa$^{-1}$. Therefore, an overburden pressure of 1 GPa
is enough to suppress CO$_{2}$ degassing from a magma containing 0.5 wt \% CO$_{2}$.
A planet with a massive ocean cannot degas. However, it can regas, provided that hydrous/carbonated minerals pass the line of arcs at subduction zones.
This suggests a steady state ocean mass over geodynamic time.

Much more pressure is needed to shut down melting than is needed to shut down degassing (Figure \ref{OVERBURDENMELT}). However, numerical simulations of water delivery during late-stage accretion \citep{ray04,ray06} produce some planets with the necessary ocean volumes to shut down melting -- the stochastic nature of late-scale accretion introduces a great deal of scatter. The largest planets we consider may accrete and retain a significant amount of nebula gas \citep{iko01}, which would also frustrate melting.

Note that even for an ocean of constant depth, land is unlikely. Gravity defeats hypsometry. For
example, doubling Earth's gravity would reduce the land area by a factor of eight. This is important for
climate-stabilizing feedback loops involving greenhouse-gas drawdown \citep{wal81}. Submarine weathering is probably less dependent on surface T than subaerial weathering,
so in the case of a water-covered planet we would expect a weaker stabilizing feedback on planet temperature from CO$_2$--silicate weathering.

\subsection{Melt-residue density inversion: decoupling melting from mantle degassing?}
Melts are more compressible than mantle minerals, so at sufficiently high pressures melt will be denser
than its residue;  for example, mid-ocean ridge basalt becomes denser than garnet
at 12.5 –-- 19.5 GPa \citep{age98}. If the sinking rate exceeds mantle velocities, these denser melts may
accumulate at the base of the mantle \citep{oht01}. Meanwhile, the residue will continue to rise to shallow pressures,
where it will eventually generate melt less dense than itself. That melt will segregate, ascend, and form
a crust. However, because atmosphere-forming volatiles are highly incompatible, they will partition
into the early-stage (sinking) melt. Because erupted melts will be volatile-poor, planets
subject to this process will be unable to balance atmospheric losses to space.
Generating melt at sufficiently high pressures for the density inversion to come into play
requires very high potential temperatures. In our modelling, these are not encountered for planets in
plate tectonic mode (except during the first 1 Gyr, which is a transient associated with cooling from our
high initial temperature; initial conditions are thus very important).
 But the density crossover is encountered for $>$ 5 {\Me}  and $<$ 3 Gyr in
stagnant lid mode. We speculate that in these planets, melting is decoupled from mantle degassing.

\section{Discussion and conclusions}

\subsection{Summary of results}

\begin{list}{$\bullet$}
{ \setlength{\labelwidth}{2cm} }
\item \emph{Modest effect of planet mass on temperature and crustal thickness:}  Scaling analysis predicts $T_{m}$ should only be weakly dependent on planet mass \citep{ste03}. Because of the exponential dependence of viscosity on temperature, modest variation in $T_{m}$ can accommodate the range of heat flows generated by planets of varying masses. Our more detailed study confirms this, and although the nonlinearity of mantle melting leads to  greater variations in crustal thickness, these are still of order unity (\S 3.2; \S 3.3; Figure \ref{CRUSTVSTIME}). However, planet mass can determine which mode of mantle convection the planet is in (\S 4), which in turn determines whether melting is possible at all (\S 3.3).
\item \emph{Challenges for plate tectonics on massive Earths:} Plate buoyancy is an increasingly severe problem for plate tectonics as planet mass increases (\S 4.5). Because Early Earth faced the same challenge, geologic fieldwork may determine the threshhold beyond which plate tectonics shuts down (\S 4.5). Middle-aged super-Earths may suffer from continental spread, which could choke off plate tectonics (\S 4.4). Shutdown of plate tectonics could place a planet in stagnant lid mode.
\item \emph{Shutdown of melting on old, stagnant lid planets:}  All our melting models predict that stagnant lid planets planets older than Earth cease melting and volcanic activity (\S 3.3-5, but see caveats in \S 6.2). Therefore, observed volcanism on rocky planets $>$ 8 Gya would provide some support for plate tectonics -- if and only if tidal heating could be shown to be small. A candidate planetary system has been identified around $\tau$ Ceti \citep{mar02}, a 10 Gya star at 3.65 parsec, and may provide an early test of this prediction.
\item \emph{Implications for the stability of atmospheres, climate, and habitability:} The atmospheres of Venus, Earth and Mars were produced by release of gases dissolved in partial melts, as well as temperature-dependent exchange of volatiles between the atmosphere and surface rocks. Planets lacking dynamos and on close orbits around their parent stars may experience high rates of atmospheric erosion due to stellar winds and coronal mass ejections \citep{kho07}, and their atmospheres would persist on main sequence timescales only if maintained by mantle melting. Under more benign conditions, crust formation may establish the long-term carbon dioxide content of Earth-like atmospheres; CO$_2$ released in volcanism is sequestered as carbonates produced during the low-temperature weathering of that crust \citep{wal81}. Volcanism may underpin the long-term habitability that the fossil record
teaches us is a prerequisite for advanced life \citep{but07}.  

 \end{list}

\subsection{Overview of approximations and model limitations}

In this paper we assume whole-mantle convection.
Layered mantle convection can alter the volcanic history of a planet
by introducing long-term sensitivity to initial thermal conditions.  For a phase boundary to form a barrier to flow its Clapeyron slope must be strongly endothermic. Historically, the $sp \to pv$ transition was thought to have this character but this is now known not to be the case. Deeper phase transitions are either exothermic ($pv \to ppv$) or very gradual \citep{sea07}. In spite of these arguments against layered mantle convection, barriers to flow are possible if stable deep mantle layers arise during or shortly after fractional crystallization of the primordial magma ocean.
Although CMB pressures reach 14 Mbar for 10 \Me and 40 Mbar for 25 \Me, 
we have not considered
metalized silicates \citep{ume06}, nor the possibility that lower-mantle convection is extremely sluggish (even isoviscous;
\citet{fow83}) due to low homologous temperatures, nor pressure dependent
viscosity \citep{pap08}. Our parameterized treatment of mantle convection
assumes the solid state and is inappropriate for very high temperatures, when the greater part of the
lithosphere is underlain by a magma ocean. For almost all cases, however, the transition to a magma ocean
takes place at temperatures beyond the range of validity of our melting models,
so magma ocean development is not the limiting consideration in interpreting our results. 

Our model does not consider tidal heating, which may be important for Earthlike planets
on close eccentric orbits about low-mass stars \citep{jac08}, nor does it take into account the
energetics of the core \citep{nim07}. Because the magnitude of core cooling is set by mantle cooling,
core energetics only dominate mantle thermal evolution in Earthlike planets as $t\to\infty$ and $H\to$ 0. Surface
temperature is assumed to be constant, which is likely to be a good approximation except for close-in
exoplanets with thin atmospheres in a 1:1 spin-orbit resonance \citep{gan08}.

All our melting models show `solidus rollover' at high P - that is, $\frac{T_{sol}}{P} \to \frac{\partial V}{\partial S}$. As a consequence, plots of $X$ versus $P$ at high $T_p$ have a long, thin high-pressure tail, and the pressure of first melting diverges at finite $T_p$. Although solidus rollover is a real effect, resulting from the greater compressibility of silicate melts versus solids \citep{ghi04}, infinite pressures of first melting are unphysical. In particular, phase transitions near 14 GPa introduce unmodelled kinks of the solidus. To sidestep the solidus rollover problem, we truncate our melting integration at 8 GPa, and do not plot model output where melting models predict nonzero $X$ at 8 GPa. But this solution has costs: 1) we cannot model hot planets (grayed-out regions in Figures \ref{CRUSTVSTIME} \& \ref{CRUSTVSTIMESL}), and 2) because pressure at first melting is so sensitive to solidus rollover, we have little confidence in our model output for the rate of processing of the mantle through the melt zone. Melting models calibrated to higher pressures and temperatures would remove this impediment to progress. One such model (xMELTS), based on a high-pressure equation-of-state
for silicate liquids \citep{ghi04}, will soon become available \citep{ghi07}. Because crustal production rate is proportional to the integral under the curve of $X$ versus $P$, rather than the pressure at which $X = 0$, it is much less sensitive to this problem. Representing the solidus as a straight line in T-P space \citep{pap08} is geologically not realistic, but has the advantage that both crustal thickness and the pressure of first melting are finite for all $T_p$. 

 We calculate crustal thickness as the integral of melting fraction from the surface to great depth. 
In plate tectonic mode this corresponds to the observable (seismically-defined) crustal thickness.
But in stagnant lid mode, the observable crustal thickness need not be the same everywhere (because melting will only occur above mantle upwellings).
Also, the observable crustal thickness may be everywhere greater than the integral of melting fraction from the surface to great depth, because crust generated in previous melting events is located directly underneath `fresh' crust. This situation is described in the context of Io by \citet{moo01}. 

On stagnant lid planets, volcanism may be episodic (as on Mars; \citet{wil01}) and absent
for long periods. This means that an observed lack of compounds with a short photochemical lifetime
would not preclude a high level of volcanic activity averaged over a sufficiently long period ($\ge 10^{\sim 8}$ yr).
Volcanism is not the only possible source of degassing. Metamorphic decarbonation \citep{bic96},
and the episodic release of deep-seated volatiles as on the Moon \citep{gor73}, could permit
(low) rates of degassing on non-volcanic worlds.

By assuming that all melt reaches the surface, we have ignored the distinction between extrusion and intrusion.
However, extrusion can alter the reflectance properties of the planet without sustaining an atmosphere -- consider the Moon -- and it may eventually be possible to discriminate between the two possible fates for melt.

\subsection{Comparison with Solar System data}

\subsubsection{Thermal evolution, with an emphasis on Earth}

Four decades since plate tectonics became widely accepted, geologists have not determined the detailed thermal history of the Earth, neither have we even produced a model that
satisfyingly accounts for the few data points in hand. The model of Korenaga (2006, taking a parameterized approach) may be an exception, but 
this model assumes that present-day observed averages are valid for the past, and is probably not portable to other Earth-like planets. Complicating the picture are two sharp-edged lower
mantle anomalies mapped by seismic tomography under Africa and the S. Pacific, which are likely distinct in composition and radiogenic-element density from the rest of the mantle, are probably not sampled in our
collections of mantle rocks brought up by volcanoes and which may be persistent, ancient features that have interfered in mantle convection since more than 2.5 Gyr ago. Their origin is unknown \citep{gar08}. Geoneutrino detectors will break some of these degeneracies before 2020, and continued field mapping of ancient tectonic features will sharpen the history of past dynamics of plate deformation \citep{con08}. But using simple models and making changes along one dimension only -- the approach taken here -- is still a meaningful approach because the more complex models introduce many more free parameters, and it is not clear how these free parameters vary with mass. By contrast, our approach involves only one `tuning' parameter, the offset $T_m - T_p$.

 The situation for the other terrestrial planets
 -- single plate planets with stagnant lithospheres - is in many ways better. Parameterized cooling fits all the observations. This is not just because of a lack of data: stagnant lid planets seem to be truly simpler, and the relevant fluid physics is probably better understood than the `viscoelastoplastic plus damage' phenomenology needed to accommodate the observed geological motions of the Earth's plates. On Earth, mantle heat loss is largely by conduction aided by hydrothermal convection at mid-ocean ridges, but mantle heat loss on stagnant lid planets should be less sensitive to the details of surface deformation.

The degree to which the mantle cools over geological time depends on the activation energy for
deformation of mantle rock, which is known only imprecisely \citep{kar08}. We take $T_{\nu} \cong$  43 K, for which $A_0$ = 7 x $10^4$ K, 
giving an order-of-magnitude viscosity change for each 100K temperature increment
 (although $T_{\nu}$ could be as high as
100K) \citep{sle07}. This is conservative (but not neccessarily more correct) in terms of the effect of mass on temperature, because small
values of $T_{\nu}$ allow mass (heat flux) variations to be accommodated by small changes in temperature
(Figure \ref{TNU}).

The range of mantle states that permit plate tectonics to continue will differ from the range of
states permitting plate tectonics to initiate, and ongoing plate tectonics will alter the state of the mantle.


Continents affect thermal evolution (\S 4.4). However, continent growth and survival is not well understood.
Hydration of
the oceanic crust at mid-ocean ridges, and subduction of this H$_2$O, is probably required for continental growth \citep{cam83}.
There is geological and isotopic evidence for episodic continental growth \citep{con06}.
Durable continents may require photosynthetic life both in the oceans \citep{ros06} and
on land \citep{aeo08}. None of these effects are straightforward to model. So, though
excessive continental surface area and radiogenic-element sequestration both have the potential to
throttle our predicted increase in mantle melting on high-mass Earths, we consider this only a tentative
prediction.

\subsubsection{Melting: constraints from Venus, Mars, Io, and the Moon}
If massive Earth-like planets
are in heat-pipe mode for $\sim$ 1 Gya (\S 4.2), a substantial fraction of the census of nearby massive Earth-like
planets would be in heat-pipe mode. And if Io is any guide (e.g., \citet{lop07}), dramatic spatial
and temporal variations in thermal emission should be present. However, the relatively slow cooling in our thermal evolution model is inappropriate for magma
oceans \citep{sle07}, because our parameterization of viscosity does not include the fifteen-order-of-magnitude
 decrease associated with melting. Planets with $Mo >
1$ are likely to cool quickly to $Mo < 1$; we suspect that Io-type cooling is unsustainable for geologically
significant periods without a nonradiogenic source of energy.

Our melting model is tuned to Earth only, but fits Venus observations reasonably well. With the MB88 model, our calculations show that
volcanism on a planet in stagnant lid mode and with Venus' mass ({\Me} = 0.85) will cease after 3.9
Gyr, which is consistent with observations \citep{sch92}.

However, when decompression melting of mantle material at background temperatures cannot
sustain volcanism on Earth-like planets undergoing whole-mantle convection, other mechanisms may take over. Plumes rising from
the core-mantle boundary, or from a compositional interface within the mantle, can have temperatures up to
several hundred degrees greater than that of passively upwelling mantle. This requires a more detailed treatment of thermal evolution than
we have taken in this paper, and could be the basis of a more detailed study.

In particular, Mars' continuing --– although fitful and low-rate –-- volcanic activity \citep{bor05}
 indicates that volcanic activity can continue on planets in stagnant-lid mode for longer than our
simple models predict, perhaps as the result of plumes in a compositionally-layered mantle \citep{wen04}
 or a thick, thermally insulating crust \citep{sch07}.
On Venus, mantle fluxing
by sinking delaminated lithosphere has been proposed as a mechanism that would allow volcanism to continue
through to the present day \citep{elk07}, although there is no robust evidence of ongoing
volcanism. Continued Solar System exploration is needed if we are to fully exploit nearby data points
to understand the geodynamic window for life.

\bigskip
\bigskip

Many of the ideas
discussed here are drawn from the conversation of Dave Stevenson, whose support for the early stages
of this project was invaluable. We thank Nick Butterfield, Rhea Workman, Brook Peterson, Norm Sleep, and Itay Halevy for their productive suggestions.
We acknowledge support from the NASA Astrobiology Institute.
E.K. acknowledges support from the Berkeley Fellowship and a Caltech Summer Undergraduate Research Fellowship.

\pagebreak




\pagebreak

\pagebreak

\begin{deluxetable}{lrrrrlccccccccccc}
\tabletypesize{\scriptsize}
\tablecaption{Radioisotope data: half lives, specific power $W$, and concentrations $[X](i)$ (ppb) after 4.5 Gyr. \label{tbl-1}}
\tablewidth{0pt}
\tablehead{
\colhead{ } &  \colhead{$^{40}$K} & \colhead{$^{232}$Th} & \colhead{$^{235}$U} & \colhead{$^{238}$U} & \colhead{Reference} &
}
\startdata
$t^{1/2}$ (Gyr) & 1.26 & 14.0 & 0.704 & 4.47 \\
Specific power (x 10$^{-5}$ W/kg) & 2.92 & 2.64 & 56.9 & 9.46 \\
\\
Concentrations: \\
`Mantle' & 36.9 & 124 & 0.22 & 30.8 & \citet{tur02} \\
`Undepleted Earth' &  30.7  & 84.1 &  0.15  & 21.0  & \citet{rin91} \\
CI chondrites  & 71.4  & 29.4 &  0.058  & 8.1  & \citet{and89} \\
EH chondrites &  147.8 &  2.8  & 5.8 &  13.0 &  \citet{new95} \\
\enddata
\tablecomments{$H_{i} = [X](i)  W_{i}$}
\end{deluxetable}


\pagebreak

\begin{deluxetable}{lcrrrlccccccccccc}
\tabletypesize{\scriptsize}
\tablecaption{Parameters used in interior and thermal models. \label{tbl-1}}
\tablewidth{0pt}
\tablehead{
\colhead{Parameter} &  \colhead{Symbol} &  \colhead{Value} & \colhead{Units} & \colhead{Reference} }
\startdata
Thermal expansivity, mantle & $\alpha$ & 3 x 10$^{-5}$  & K$^{-1}$ & 1\\
Thermal conductivity & $k$ &  4.18  & W m$^{-1}$ K$^{-1}$ & 1\\
 & $\beta$ &  0.3 & & 1\\
Thermal diffusivity & $\kappa$ &   10$^{-6}$ &  m$^2$ s$^{-1}$ & 1 \\
Critical Raleigh number & $Ra_{cr}$  & 1100 & & 1\\
Gas constant & $R$ &  8.31 &  J K$^{-1}$ mol$^{-1}$ & 2 \\
Specific heat capacity, mantle & $c$ &  914 &  J K$^{-1}$ kg$^{-1}$ & 1\\
Density, mantle & $\rho_{mantle}$  & 3400  & Kg m$^{-3}$ & 1\\
Density, crust & $\rho_{crust}$ & 2860 & Kg m$^{-3}$ & 3\\
Reference viscosity & $\nu_{0}$ &165 &  m$^2$ s$^{-1}$ & 1\\
Gravitational constant & $G$ &  6.67 x 10$^{-11}$ &  m$^3$ kg$^{-1}$ s$^{-2}$ & 2\\
Core mass fraction & $f_{core}$  & 0.325 & & 2\\
Earth radius & R$_{\earth}$ &  6.372 x 10$^6$  & m & 2\\
Earth mass & M$_{\earth}$ &  5.9742 x 10$^{24}$ & kg\\
Mantle temperature, initial & $T_m$ (t = 0) &   3273  & K \\
Temperature change causing e-folding in viscosity & $T_{\nu}$ &  43 or 100  & K & 1 or 4\\
\enddata
\tablecomments{1. \citet{tur02} 2. \citet{dep01} 3. \citet{car90} 4. \citet{sle07}}
\end{deluxetable}

\pagebreak
\newpage

\begin{figure}[p]
\includegraphics[width=1.0\textwidth, clip=true, trim = 60mm 20mm 60mm 240mm]{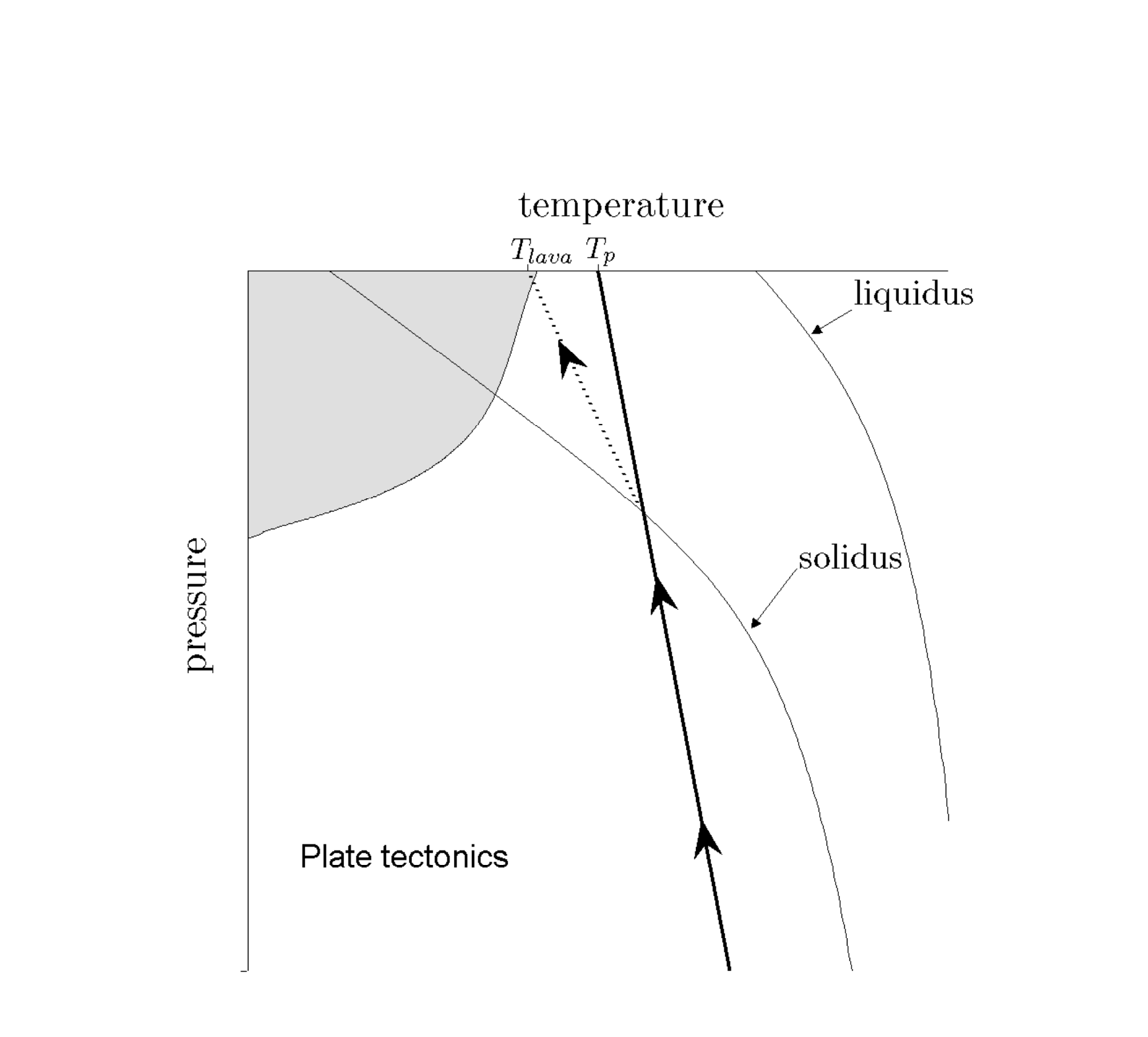}
\end{figure}

\pagebreak
\clearpage
\newpage

\begin{figure}[p]
\includegraphics[width=1.0\textwidth, clip=true, trim = 60mm 20mm 60mm 240mm]{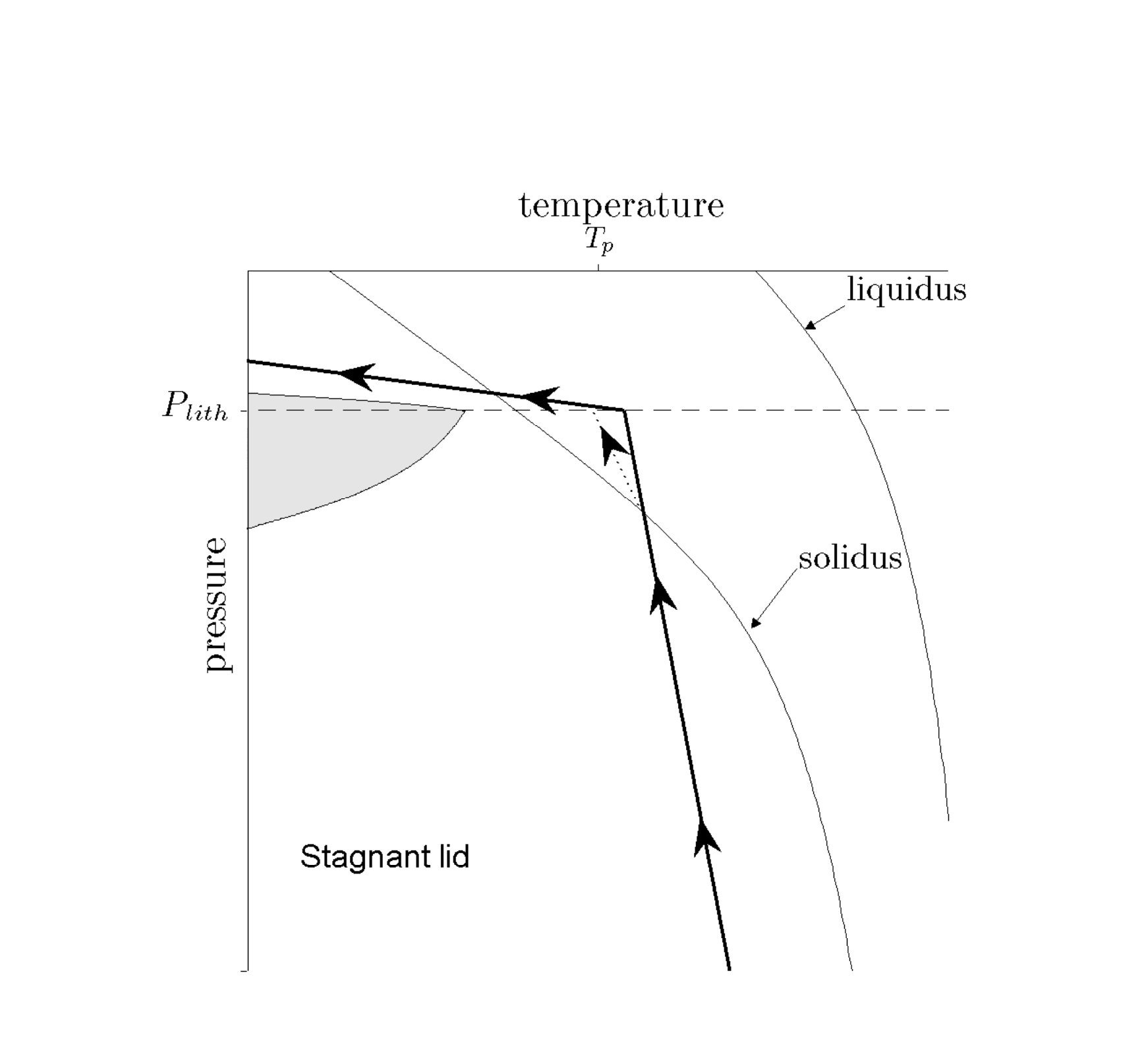}
\caption{\label{PEDAGOG} Sketches of pressure-temperature paths for passively upwelling mantle, and resulting melt fraction.
In each sketch, the shaded area corresponds to the partial melt fraction as a function of pressure.
a) Plate tectonics. Thick solid line is the adiabatic decompression path for solid mantle. Actual path taken by upwelling mantle,
traced by arrows, differs above the solidus because of latent heat of fusion.
b) Effect of a stagnant lid, whose base corresponds to the dashed line. Ascending mantle tracks the conductive geotherm within the lid (arrowed path).
Melt generated at $P < P_{lith}$ in the stagnant lid case is a small fraction of total melt, and we ignore it in this paper.}
\end{figure}
%



\pagebreak
\newpage

\begin{figure}[p]
\includegraphics[width=0.75\textwidth, clip=true, trim = 0mm 50mm 00mm 60mm]{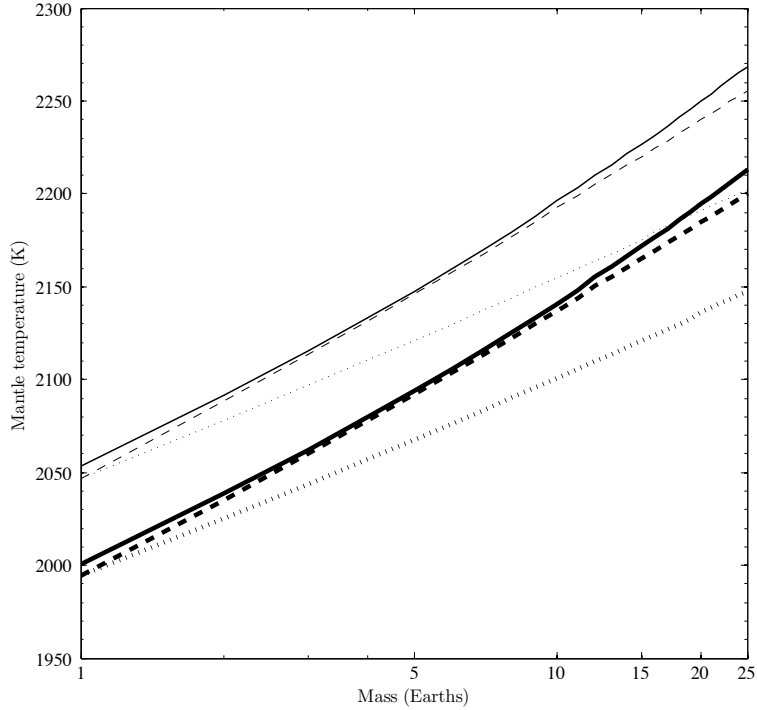}
\caption{\label{UREYEQUALS1}Effect of mass on mantle temperature for a planet in thermal equilibrium with a specific radiogenic power appropriate for today's Earth. Thick lines correspond to a surface
temperature of 273 K, thin lines correspond to a surface temperature of 647K. The solid line uses the
scaling of Seager et al. (2007), and the dashed lines use the scaling of Valencia et al. (2006): note that
the latter is only valid for M $<$ 10 Earths. The dotted lines use constant-density scaling. Seager et al. (2007) model cold exoplanets, leading to an underestimate of Earth's radius is 3 \%. The omission of thermal expansion leads to a smaller surface/area volume ratio than the other models, so the Seager et al. (2007) curves plot above those for Valencia et al. (2006) at low mass.}
\end{figure}



\begin{figure}[p]
\includegraphics[width=1.0\textwidth, clip=true, trim = 0mm 40mm 0mm 40mm]{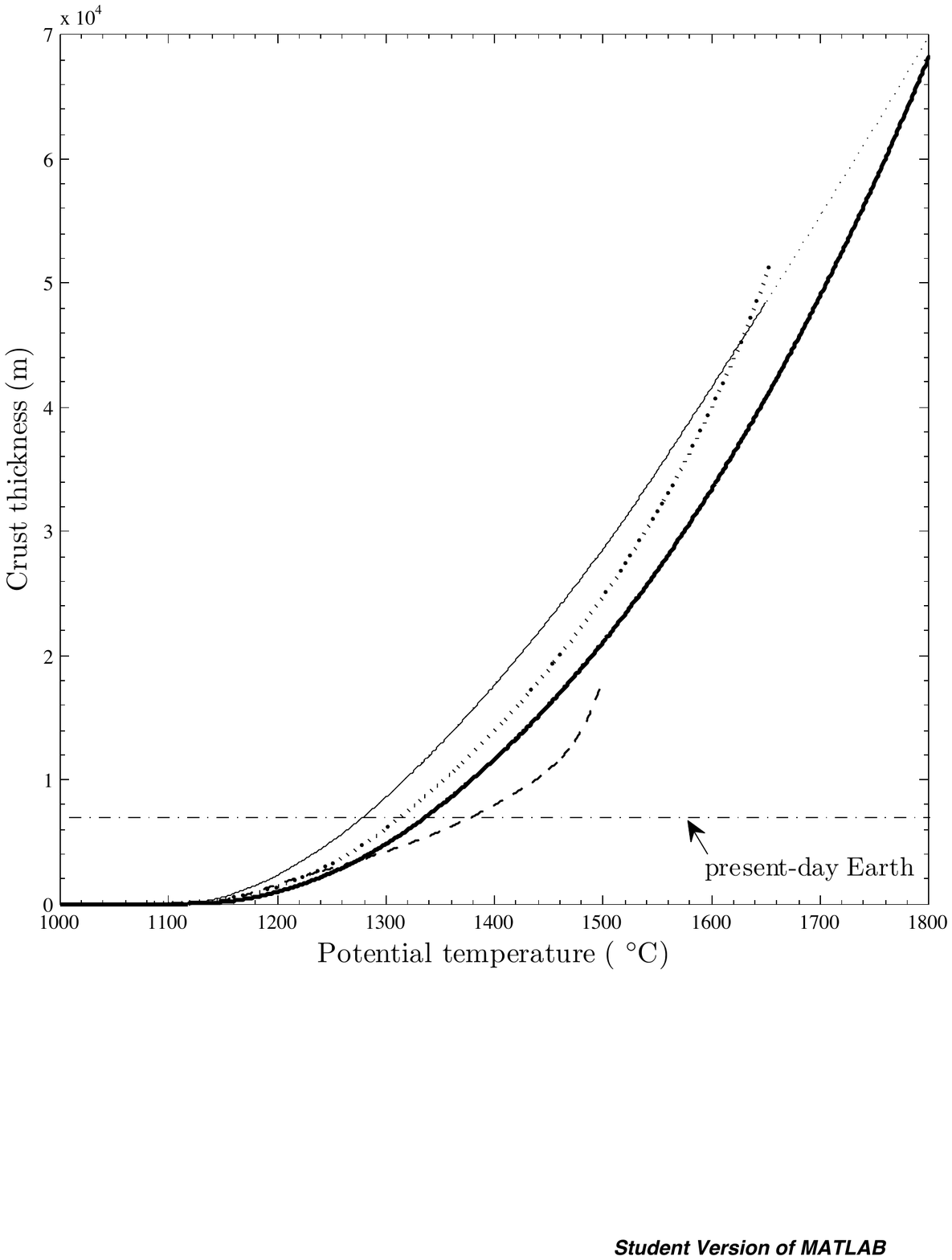}
\caption{\label{CT}Crustal thickness as a function of potential temperature for {\Me} = 1. Thick line corresponds to
the model of \citet{lan92}, which has a similar functional form to the models used by \citet{sle07} and \citet{pap08}. Thin solid line corresponds  to MB88 model, thick dotted line to K03 model, and dashed line to pMELTS model. Thin dotted line is a cubic extrapolation of MB88 beyond its range of validity. pMELTS results are only shown where first
melt occurs at $<$ 3 GPa. MB88 results are only shown where melt fraction is zero at 8 GPa. Horizontal
dash-dot line is observed crustal thickness on today's Earth.}
\end{figure}

\pagebreak
\newpage
\clearpage

\begin{figure}[p]
\includegraphics[width=0.75\textwidth, clip=true, trim = 60mm 0mm 80mm 250mm]{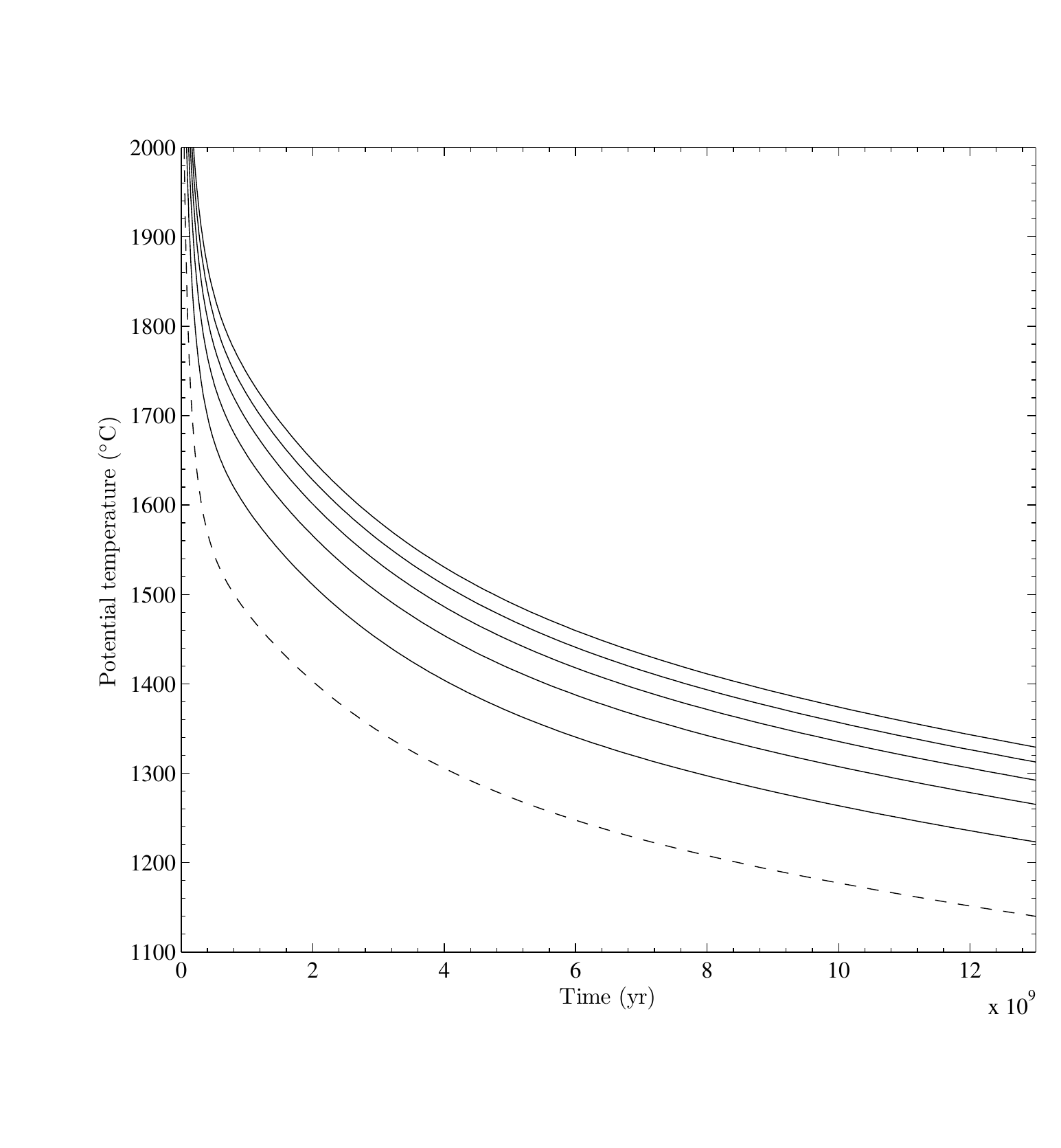}
\caption{\label{MASS}Effect of increasing planet mass on thermal evolution. Mantle temperatures are adjusted
to produce 7 km thick crust with plate tectonics under MB88 melting model at 4.5 Gyr for 1 \Me. Dashed line is 1 \Me; solid lines are
for 5, 10, 15, 20 and 25 \Me, with temperature increasing with increasing mass. T$_{\nu}$ = 43 K.}
\end{figure}


\begin{figure}[p]
\includegraphics[width=0.8\textwidth, clip=true, trim = 70mm 50mm 70mm 35mm]{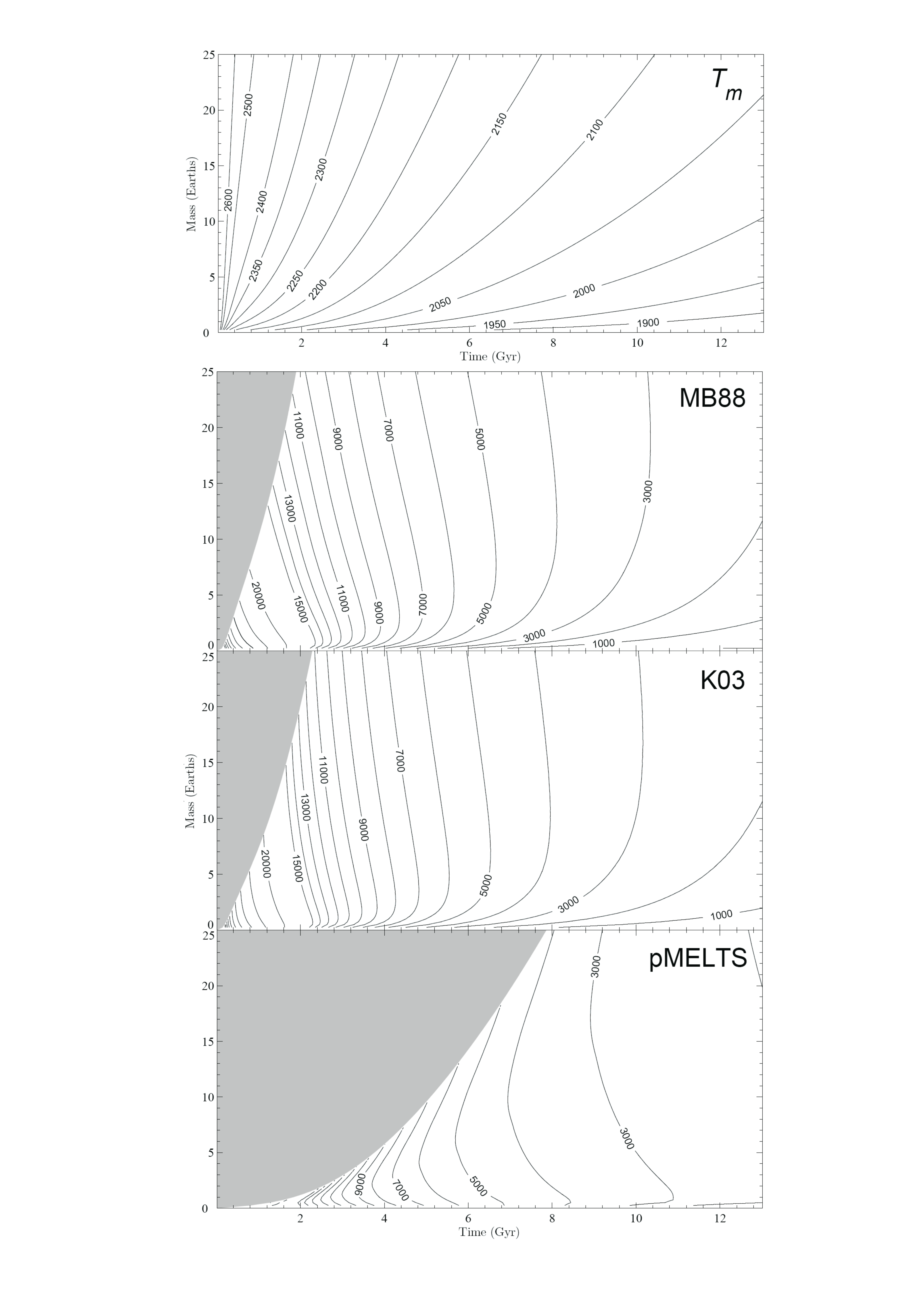}
\end{figure}

\pagebreak
\newpage
\clearpage

\begin{figure*}[p]
\caption{\label{CRUSTVSTIME}(Uppermost panel) Evolution of mantle temperature (degrees C) with time under plate tectonics. Seager et al. (2007) internal structure model.
Initial temperature for all models is 3273 K. (Lower panel, top to bottom) Corresponding crustal thicknesses in meters for MB88, K03 and pMELTS melting
modes. Contour
interval is 1000m from 0 to 15000m, and 5000m for larger values. Light grey regions are where
melting models are extrapolated beyond their stated range of validity. Horizontal line at bottom right of MB88 pane corresponds to cessation of volcanism on 0.25 \Me planets after $\sim$ 12 Gyr (see text).}
\end{figure*}

\pagebreak
\newpage
\clearpage

\begin{figure}[p]
\includegraphics[width=0.75\textwidth, clip=true, trim = 80mm 20mm 80mm 225mm]{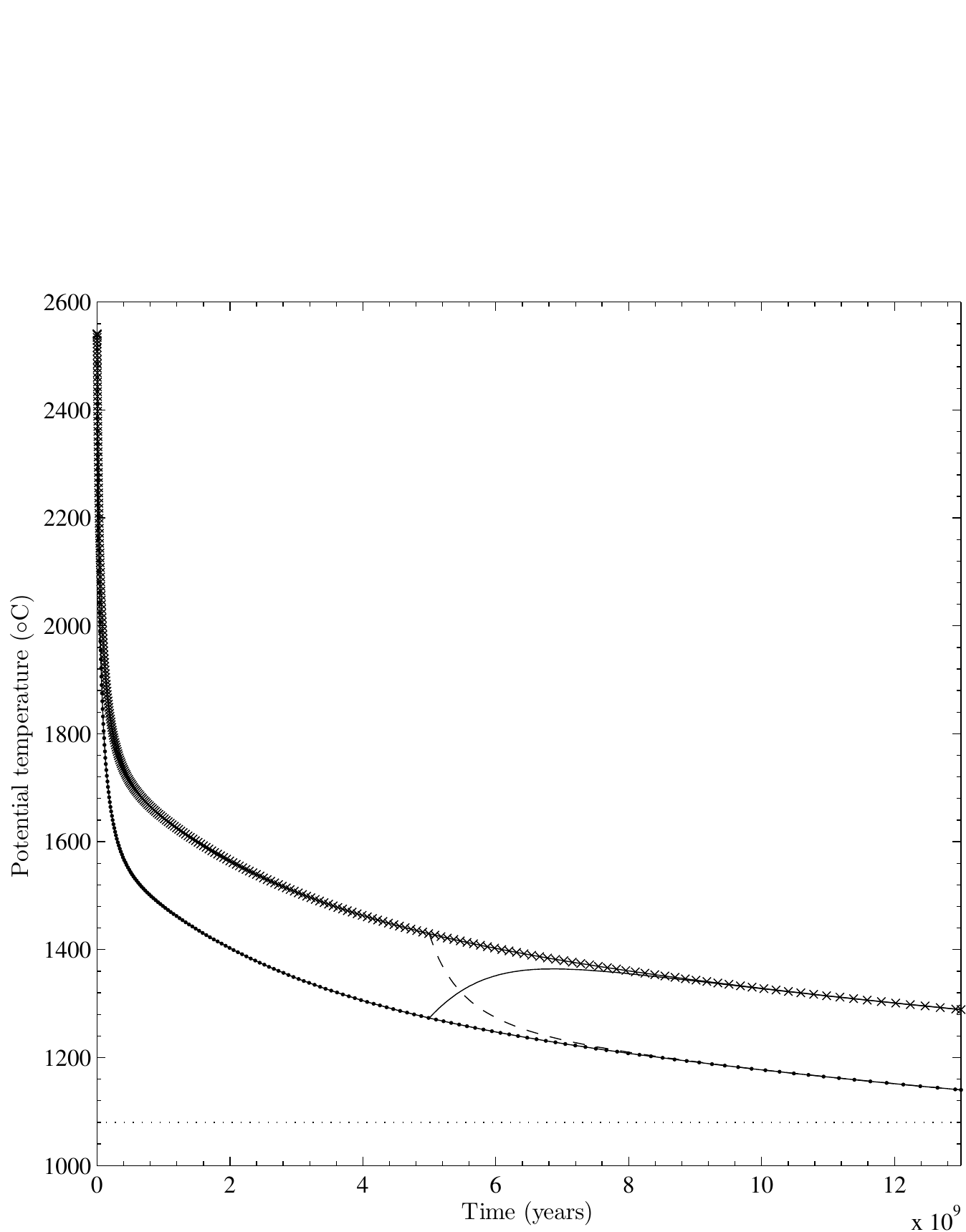}
\caption{\label{MANTLEMODE}Effect of mode of mantle convection on thermal evolution. Stars joined by solid line correspond
to stagnant lid mode. Dots joined by solid line correspond to plate tectonics mode. Thin solid line shows thermal
evolution when an instantaneous switch to stagnant lid mode is imposed, after 5 Gyr, on a planet undergoing plate
tectonics. The dotted line corresponds to the thermal evolution following an instantaneous switch to plate tectonics.
For equivalent radiogenic complements, a planet in plate tectonics mode will have a lower potential
temperature than a planet in stagnant lid mode. The difference is comparable to the range in potential
temperatures due to mass.}
\end{figure}

\pagebreak
\newpage



\begin{figure}[p]
\includegraphics[width=1.0\linewidth, clip=true, trim = 70mm 150mm 70mm 200mm]{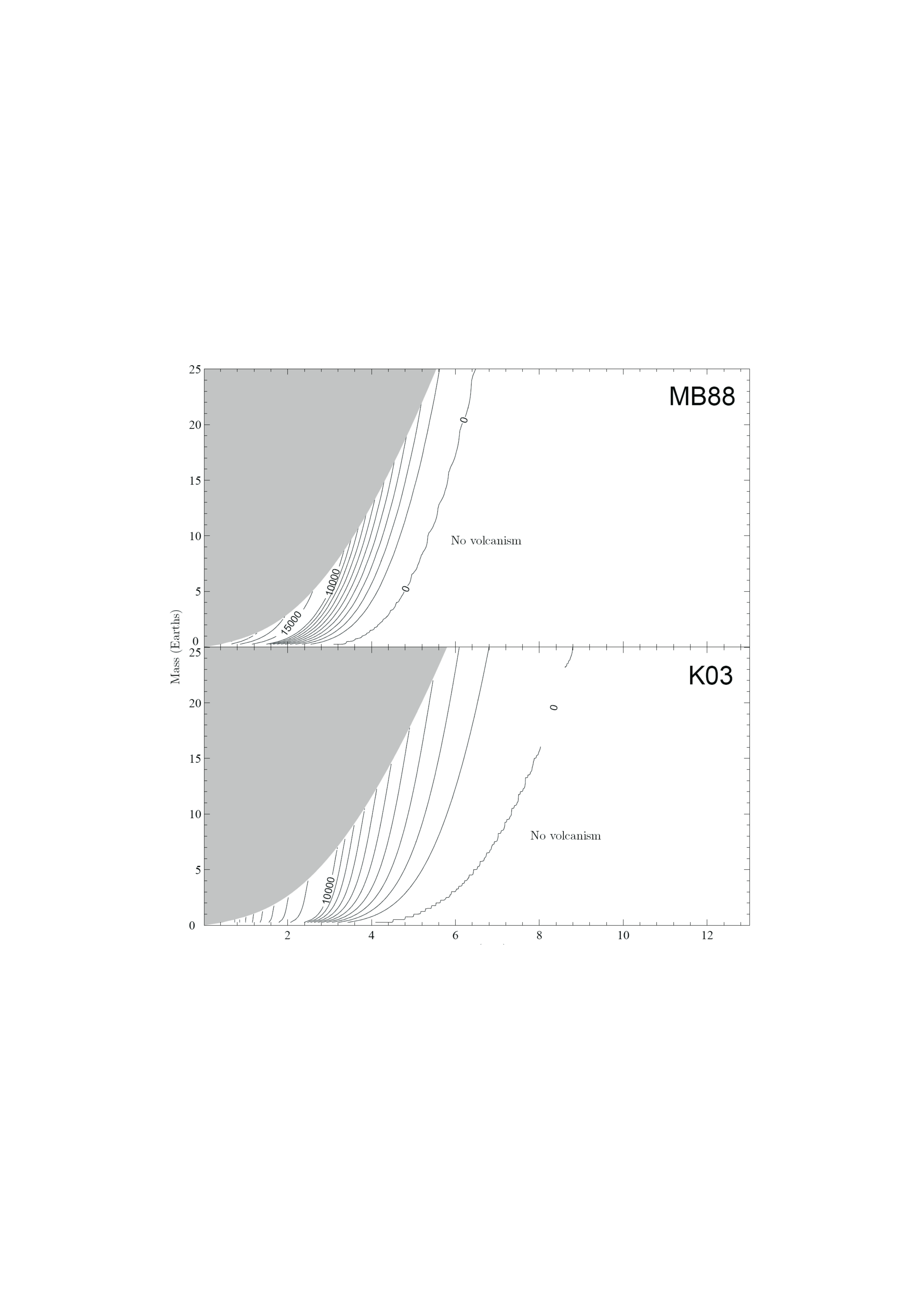}
\caption{\label{CRUSTVSTIMESL}Evolution of crustal thickness with time in stagnant lid mode, for MB88 (top)
and K03 (bottom) melting models. Wiggles in the CT = 0 contour are interpolation artifacts.}
\end{figure}
%



\clearpage
\newpage
\pagebreak

\begin{figure}[p]
\includegraphics[width=0.75\textwidth, clip=true, trim = 0mm 50mm 10mm 60mm]{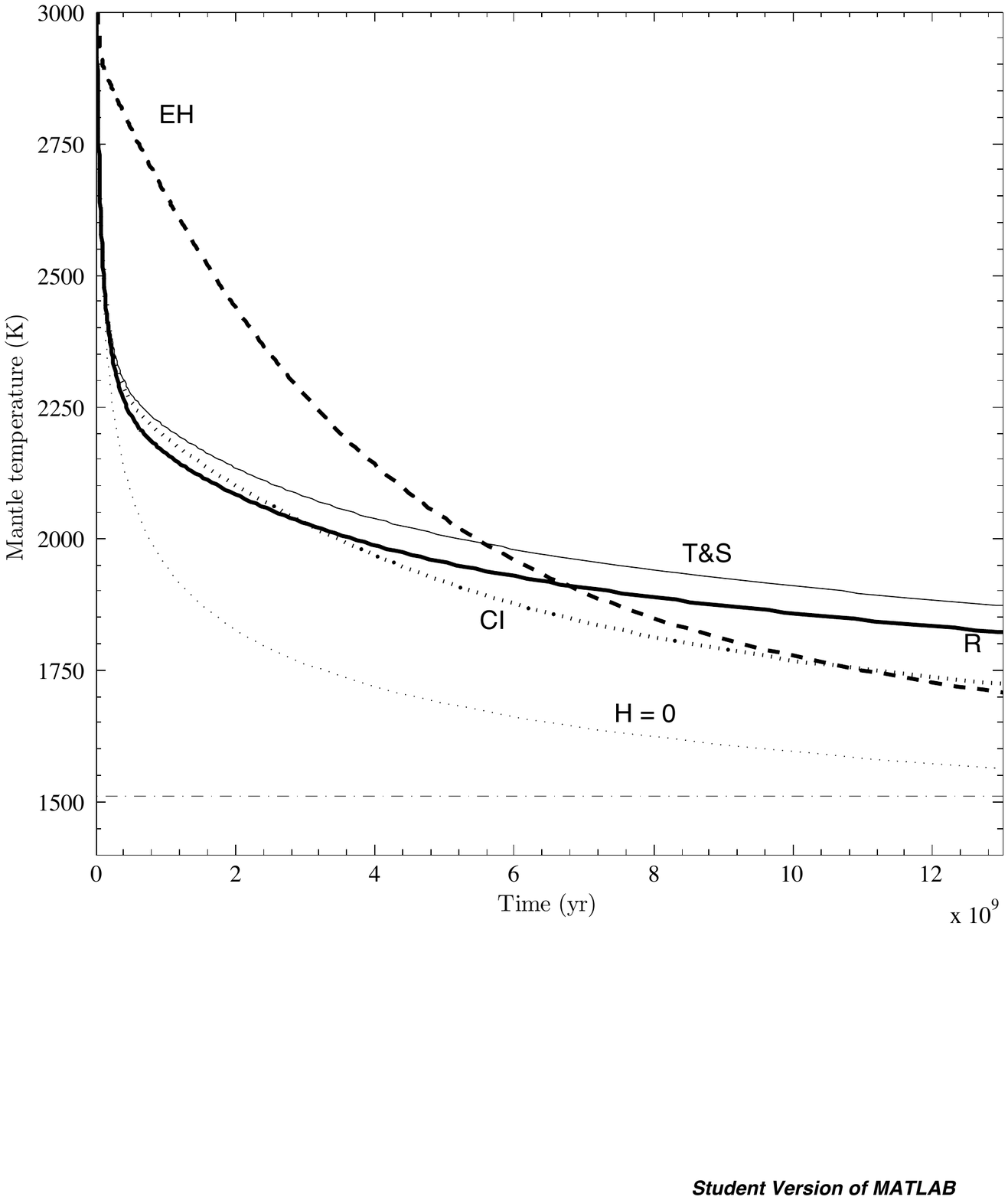}
\caption{\label{H}Effect of different radiogenic-element complements on thermal evolution. Initial mantle
temperatures are identical; all runs are with T$_{\nu}$ = 43 K. Thick lines correspond to various chondritic
scenarios: thick dashed line is EH chondrite; thick dotted line: CI chondrite; thick solid line:
Ringwood, 1991. The thin solid line is for \citet{tur02}, and the thin dotted line is for no
radiogenic elements in the mantle. This could correspond, for example, to early and complete differentiation of the mantle to
produce a thick crust, which is then swiftly removed by impacts. The horizontal dash-dot line at 1510 K corresponds to $Nu$ = 1 (no convection, no large-scale mantle flow, and no potential for sustained melt production).}
\end{figure}

\pagebreak
\clearpage
\newpage
\begin{figure}[p]
\includegraphics[width=1.0\textwidth, clip=true, trim = 0mm 50mm 0mm 50mm]{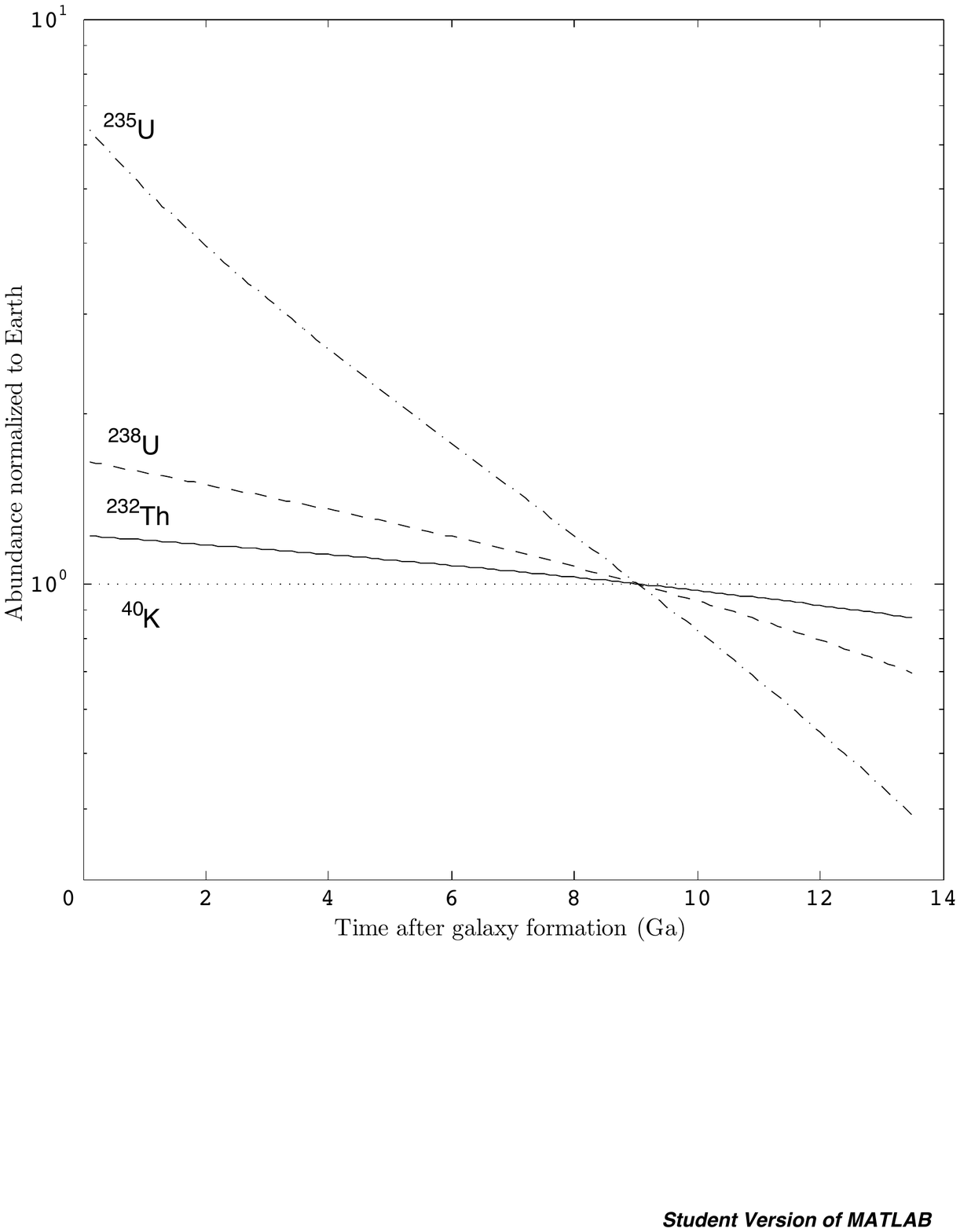}
\caption{\label{ISOTOP}Abundance, relative to silicon and normalized to conditions at the time of the protosolar nebula,
 of the principal long-lived radionuclides in rocky planet mantles.}
\end{figure}

\pagebreak
\newpage
\clearpage

\begin{figure}[p]
\includegraphics[width=1.0\textwidth, clip=true, trim = 0mm 50mm 0mm 50mm]{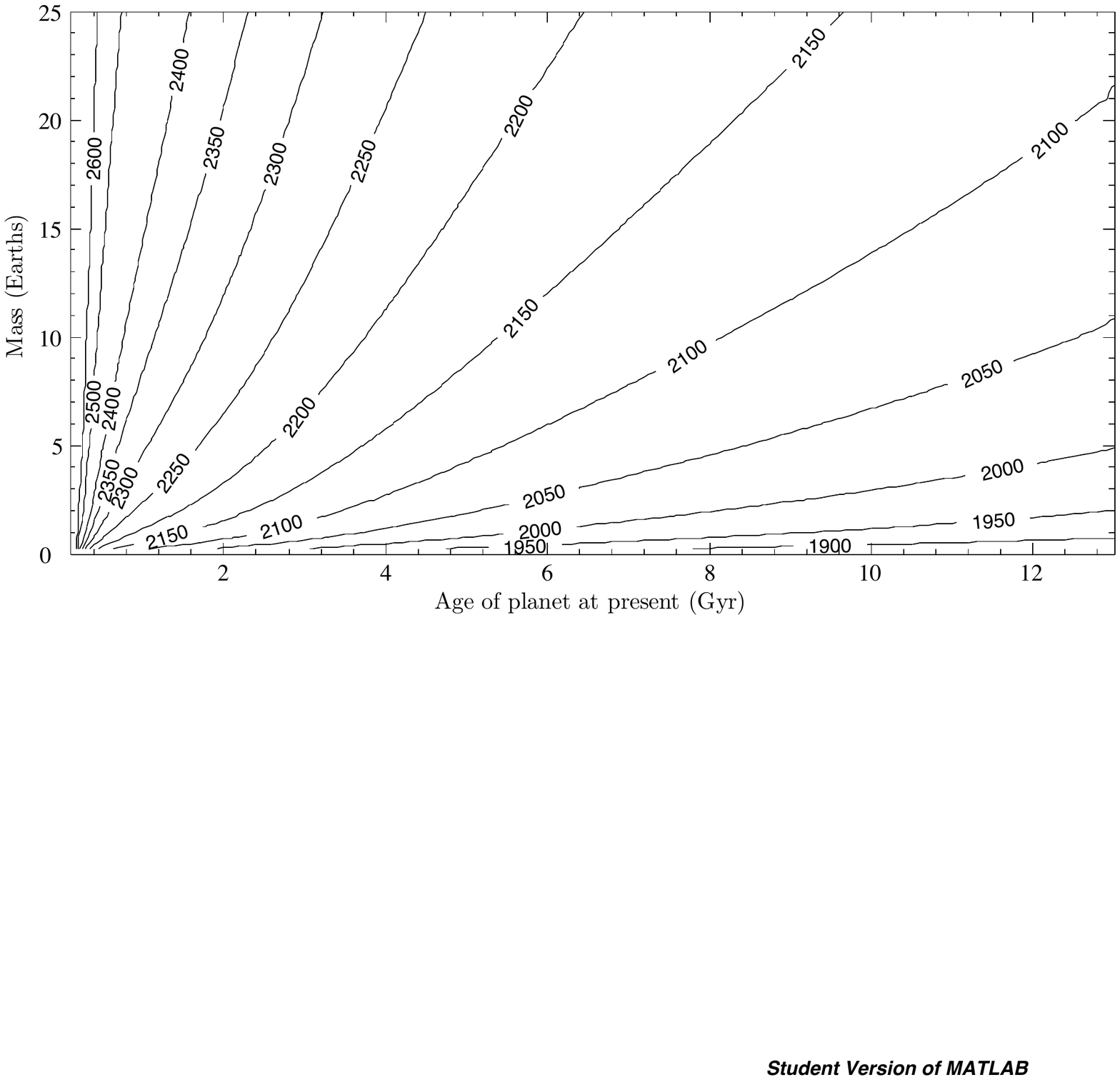}
\caption{\label{GCE}Temperature at the present epoch, tracking the effect of galactic cosmochemical evolution on initial radioisotope complement. Planets plotting to the left orbit young stars, planets plotting to the right orbit old stars. Compare with top panel of Figure 10. Note that the abscissa is not time, but the age of planet at the present day.}
\end{figure}

\begin{figure}[p]
\includegraphics[width=1.0\textwidth, clip=true, trim = 0mm 50mm 0mm 50mm]{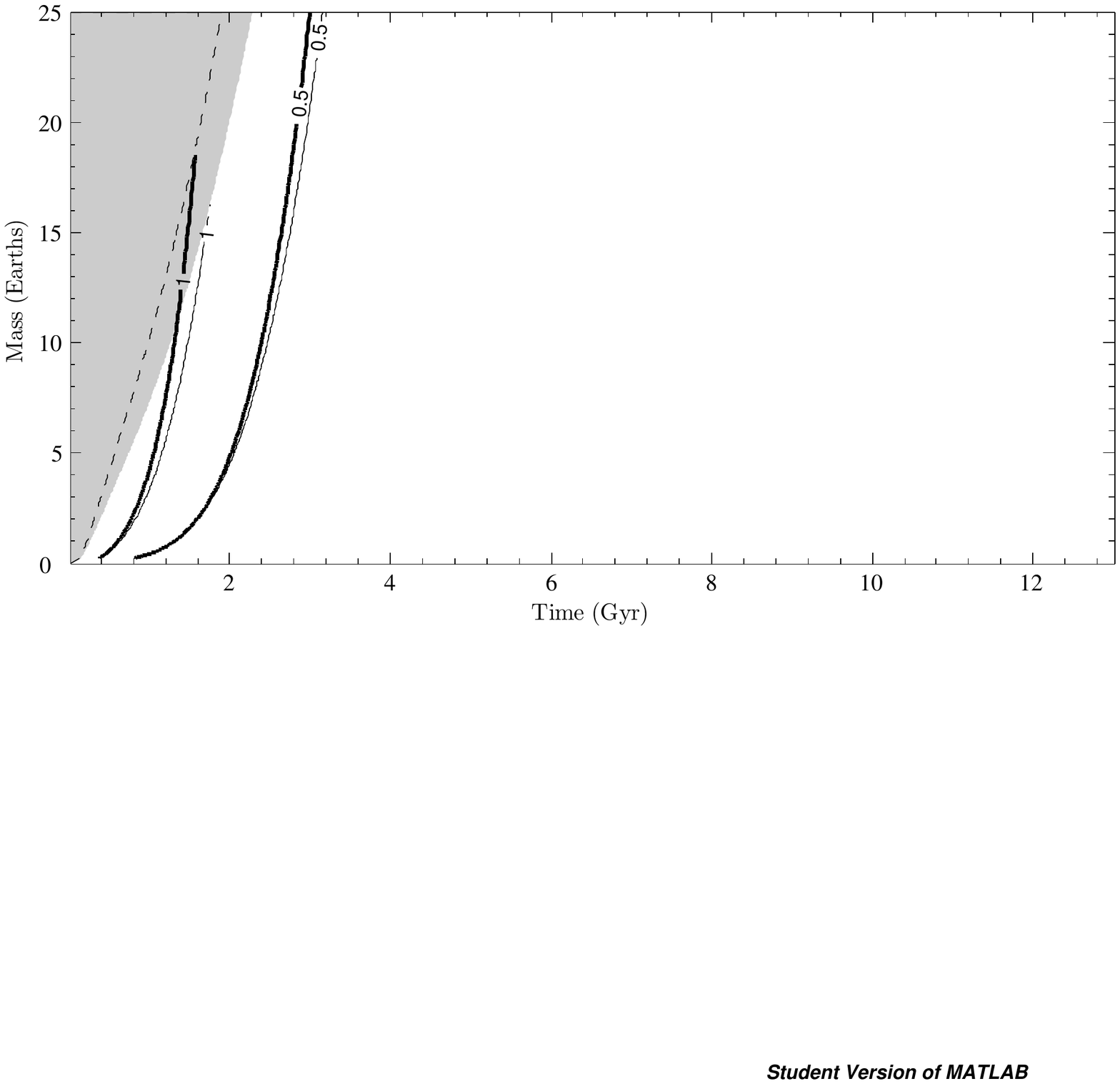}
\caption{\label{COVERL}Possible limits to plate tectonics. Thick solid lines correspond to crust--to--lithosphere
thickness ratios of 0.5 and 1.0 for MB88 melting model; thin solid lines are the same, but for K03 melting model.
Points to the left of these solid lines may be subject to vertical (Io-type)
tectonics. Dotted line is the limit of validity of the MB88 melting model.
 Results not shown for pMELTS melting model because solid lines fall in the temperature region
for which pMELTS is not valid.}
\end{figure}

\pagebreak
\begin{figure}[p]
\includegraphics[width=1.0\textwidth, clip=true, trim = 40mm 0mm 40mm 150mm]{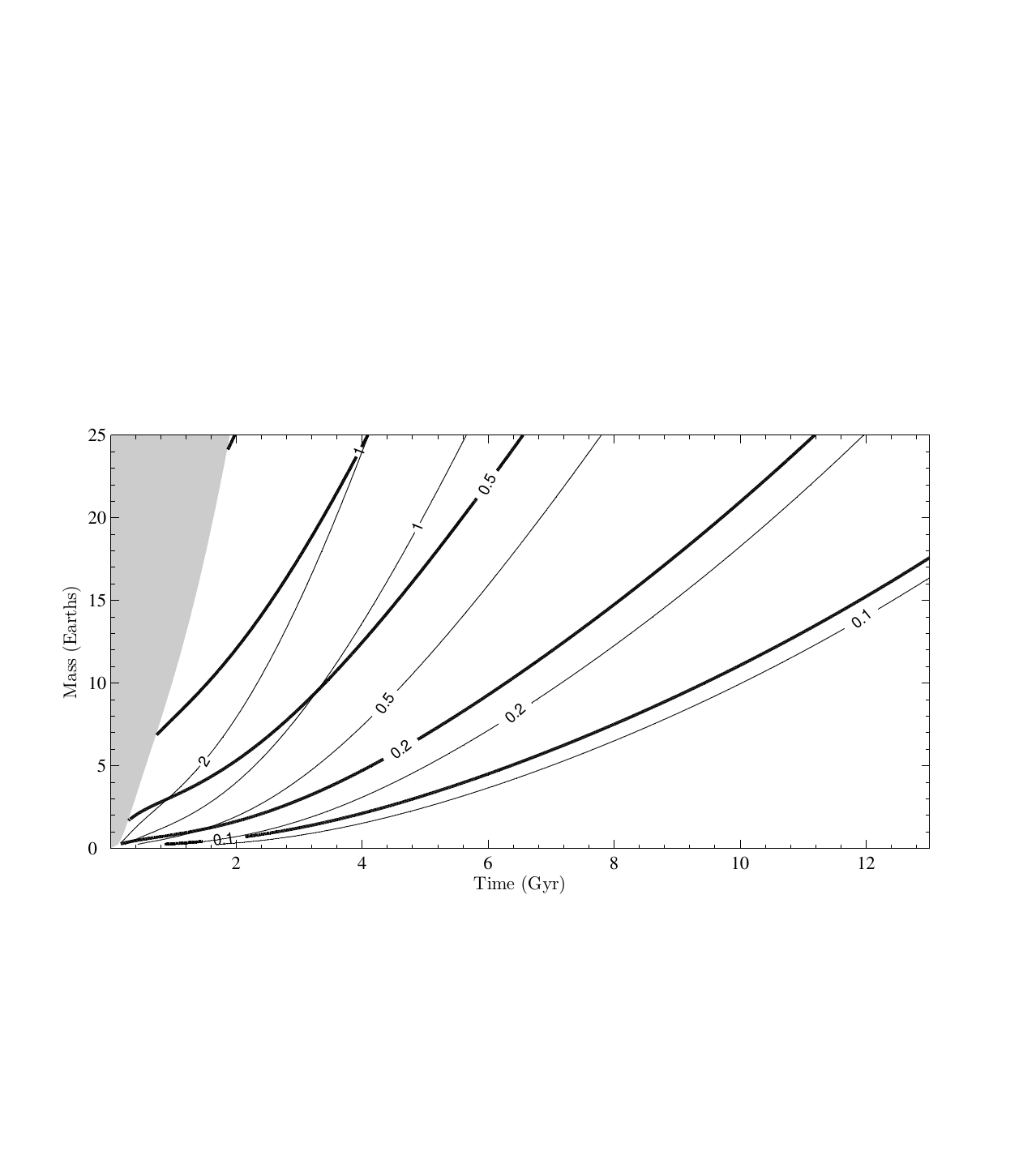}
\caption{\label{SPREADING}Plate spreading rate (m/yr) as a function of time. Thick lines correspond to
solution including heat transport through magma pipes. Thin lines correspond to conduction-only
solution. MB88 melting model.}
\end{figure}

\pagebreak
\clearpage
\newpage

\begin{figure}[p]
\includegraphics[width=1.0\textwidth, clip=true, trim = 50mm 00mm 50mm 250mm]{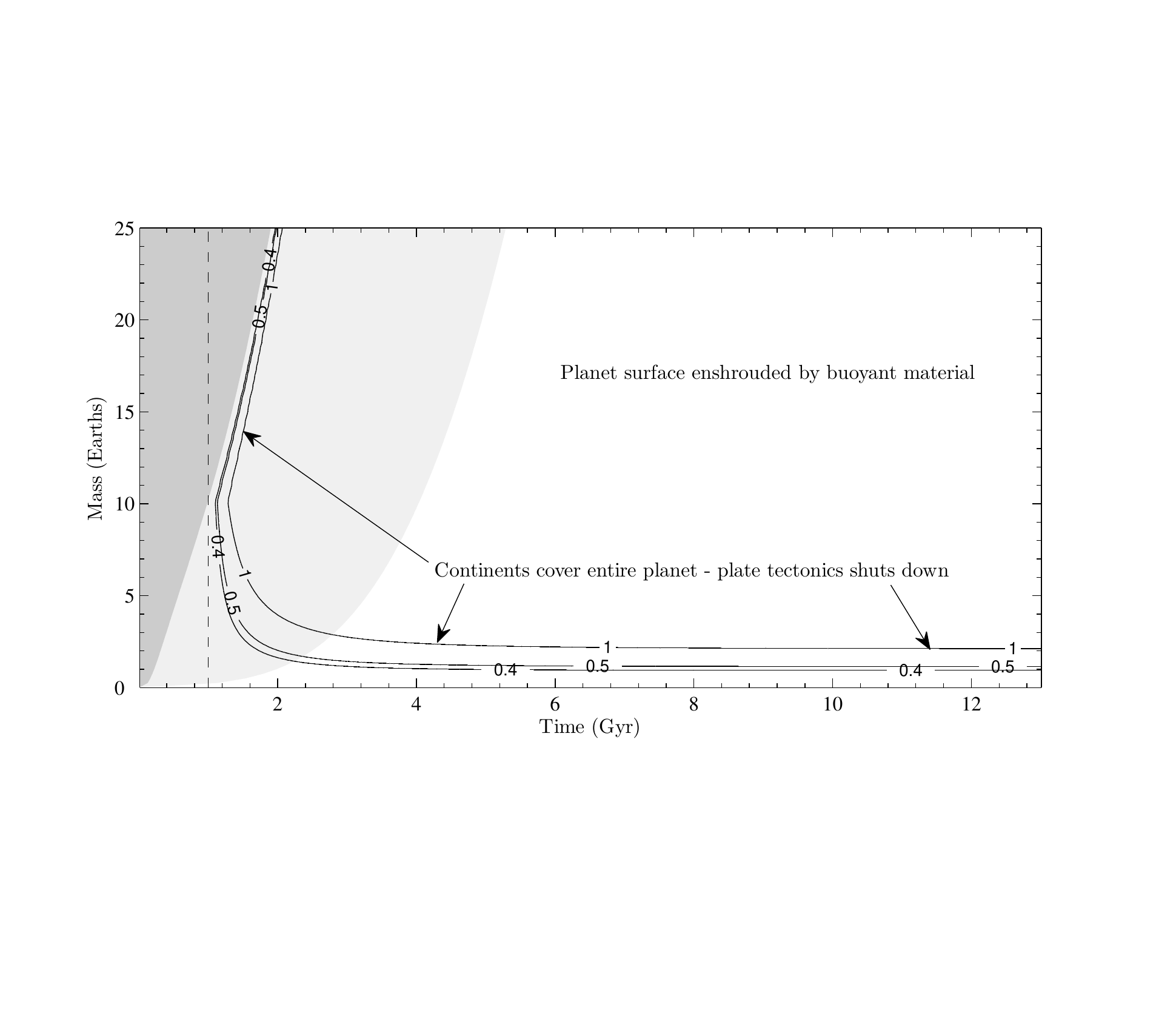}
\caption{\label{CONTTHROTTLE}Fractional area covered by continents, versus time. MB88 melting model.
The dark shaded region is that for which the melting model is invalid. The light shaded region is that for which buoyancy stresses
probably prevent plate tectonics. The vertical dashed line is the time after which continental growth is permitted (1 Gyr; \citet{con08}). For \Me $>$ 10, continents choke plate tectonics $\le$ 0.3 Gyr after continent growth is permitted.}
\end{figure}

\pagebreak
\clearpage
\newpage
%

\clearpage
\begin{figure}[p]
\includegraphics[width=1.0\textwidth, clip=true, trim = 0mm 50mm 0mm 50mm]{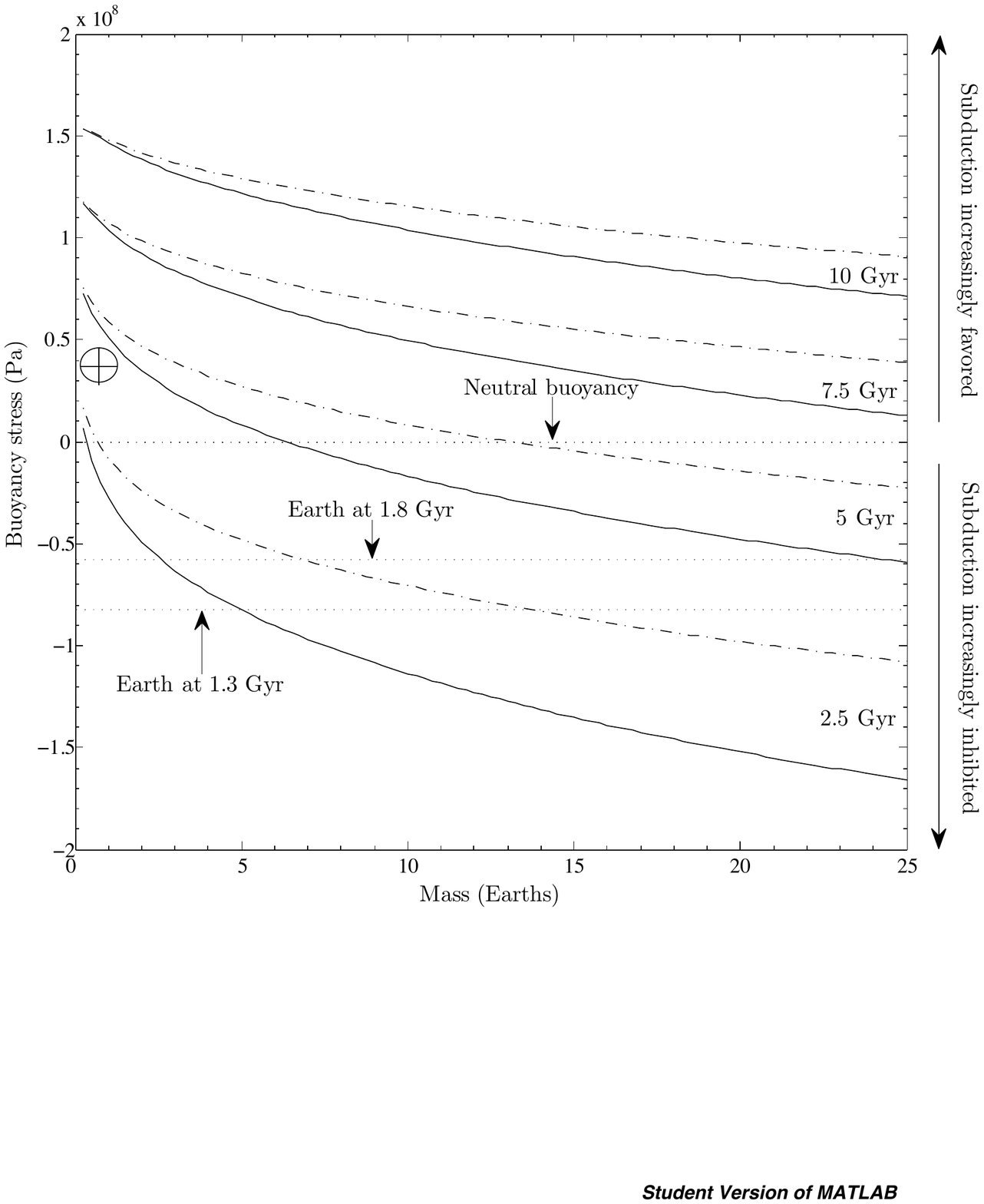}
\caption{\label{BUOYSTRESS} Buoyancy stresses as a function of thermal evolution and planet mass. Buoyancy stress is the product of density contrast, gravity, and lithospheric thickness. Positive values
denote plate denser than underlying mantle, favoring subduction; negative values denote plate more
dense than underlying mantle, retarding subduction. Solid lines connect buoyancy values for planets of
different masses 2.5 Gyr, 5 Gyr, 7.5 Gyr and 10 Gyr after planet formation, for constant crustal density
of 2860 kg m$^{-3}$. Dash-dot lines are for a crustal density of 3000 kg m$^{-3}$, as might be the case for partial
amphibolitization. Dotted lines are possible lower limits to plate tectonics based on Earth's (disputed)
geological record; arguably, subduction must be possible on planets whose buoyancy forces plot above these
lines. The Earth symbol is the model calculation for present day conditions on Earth. }
\end{figure}

\clearpage
\pagebreak
\newpage



\clearpage
\newpage
\pagebreak

\begin{figure}[p]
\includegraphics[width=0.75\textwidth, clip=true, trim = 0mm 25mm 0mm 0mm]{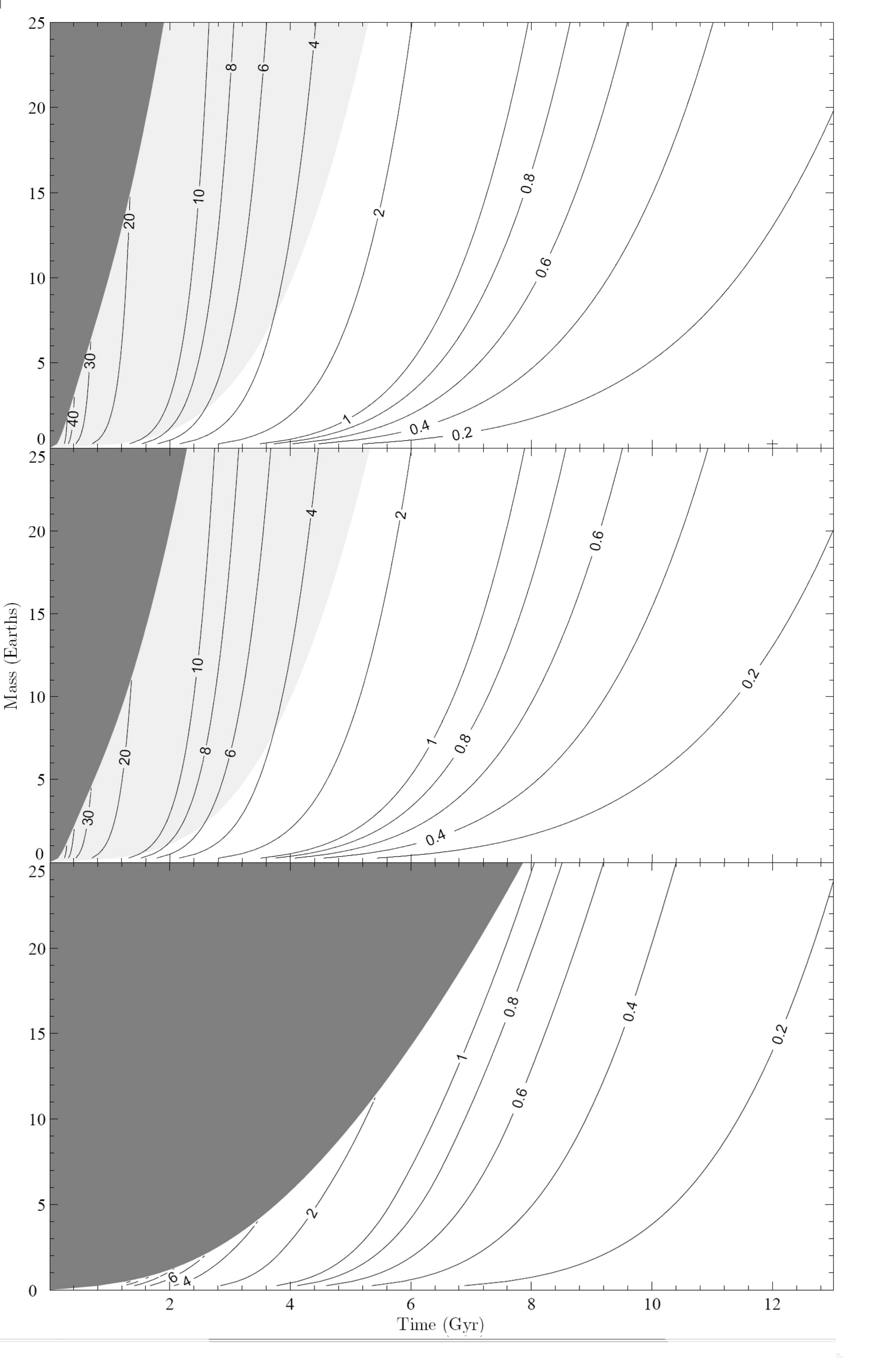}
\caption{\label{RATE}Rate of volcanism per unit mass on massive Earth-like planets experiencing plate tectonics, normalized to calculated
rate on the Earth (3.7516 x 10$^{-19}$ s$^{-1}$, equivalent to 24 km$^3$ yr$^{-1}$), for MB88 (top), K03 (middle) and pMELTS (bottom) melting
models. Light grey shaded regions correspond to negative buoyancy stresses
with magnitudes in excess of 50 MPa, which would markedly inhibit subduction. Dark grey shaded
regions correspond to mantle temperatures too high for a reliable crustal thickness calculation.}
\end{figure}

\clearpage
\newpage
\pagebreak

\begin{figure}[p]
\includegraphics[width=1.0\textwidth, clip=true, trim = 70mm 250mm 70mm 150mm]{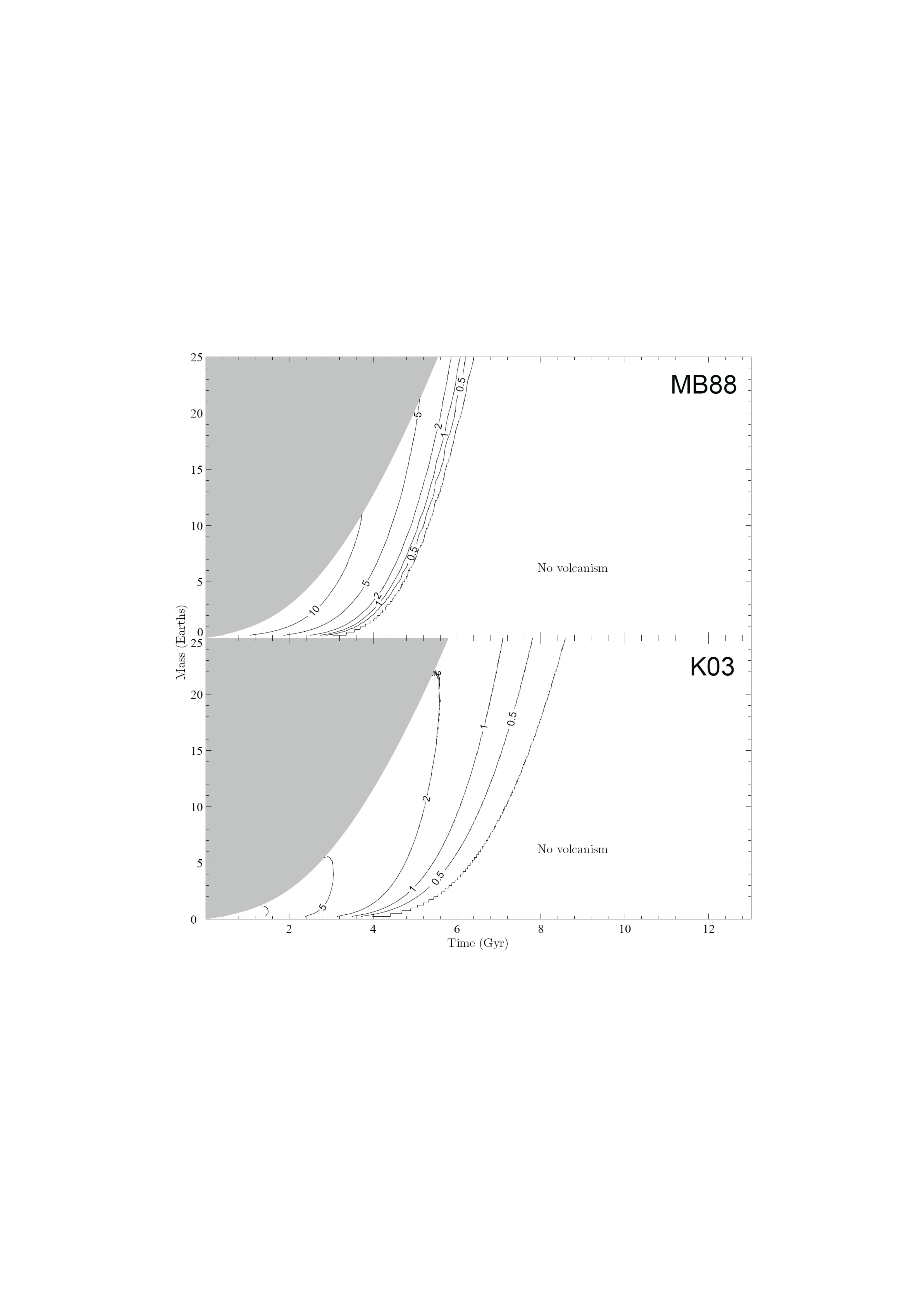}
\caption{\label{RATESL}Rate of volcanism on massive Earth-like planets undergoing stagnant lid convection, normalized to calculated rate on a plate-tectonic Earth, for MB88 (top) and K03 (bottom) melting models. Dark grey shaded
regions correspond to mantle temperatures too high for a reliable crustal thickness calculation. Contours are at 0, 0.5, 1, 2, 5 and 10 times Earth's rate.}
\end{figure}

\clearpage
\newpage
\pagebreak

%


%

\pagebreak
\begin{figure}[p]
\includegraphics[width=1.0\textwidth, clip=true, trim = 40mm 0mm 40mm 250mm]{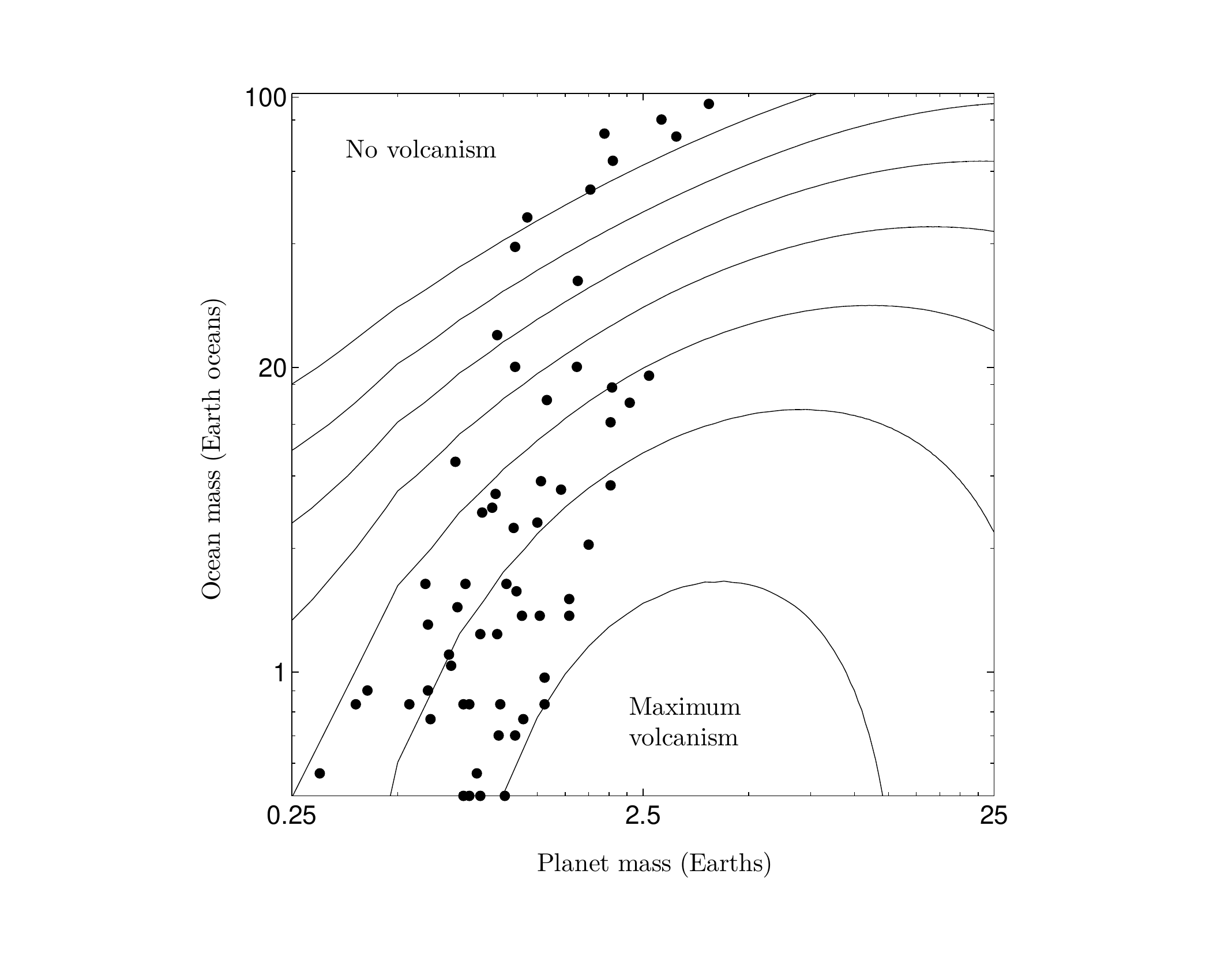}
\caption{\label{OVERBURDENMELT}Crustal thickness as a function of planet mass and ocean mass for planets with plate tectonics, after 4.5 Gyr. K03 melting model. 
Solid lines are contours of crustal thickness at 1km intervals. Small fluctuations in the contours are interpolation artifacts.
Black circles correspond to simulated planets in habitable zone from \citet{ray04,ray06},
assuming (following \citet{ray06}) that volatiles are
partitioned between surface and mantle reservoirs in the same proportions as on Earth.
 We treat the effect of the volatile overburden on melting as being equivalent to that of a
stagnant lid with the same basal pressure.}
\end{figure}

\begin{figure}[p]
\includegraphics[width=1.0\textwidth, clip=true, trim = 20mm 20mm 0mm 40mm]{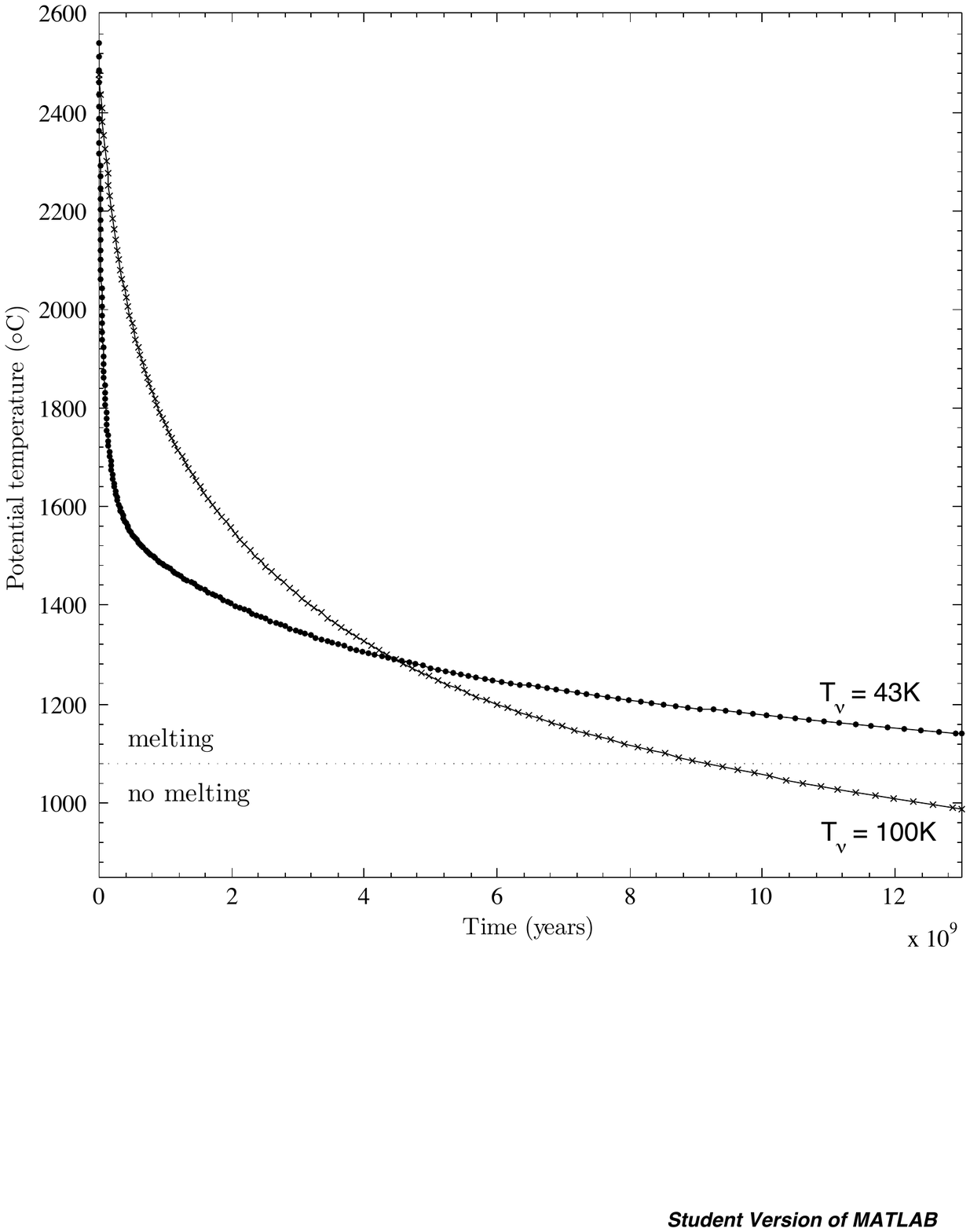}
\caption{\label{TNU}Effect of changing activation energy for temperature-dependent viscosity on thermal
evolution. \citet{tur02} radiogenic-element complement (\Me = 1). Mantle temperatures are
adjusted to produce 7 km  thick crust under MB88 at 4.5 Gyr. Dots correspond to results with T$_{\nu}$ =
43K. Stars correspond to results with T$_{\nu}$ = 100K. Note that the cooling rate at 4.5 Gyr is 71 K/Gyr for
T$_{\nu}$ = 100K, more than double that for T$_{\nu}$ = 43 K (33 K/Gyr). Cessation of melting (dashed line)
occurs around 9 Gyr for T$_{\nu}$ = 100K, but takes longer than the age of the Universe for T$_{\nu}$ = 43K.}
\end{figure}

\newpage
\pagebreak
\clearpage


\end{document}